\documentclass[%
aps,
pra,%
amsmath,amssymb,showpacs,
reprint,%
floatfix,
twocolumn,
nofootinbib,
longbibliography,
superscriptaddress
]{revtex4-1}

\usepackage{graphicx}
\usepackage{dcolumn}
\usepackage{bm}
\usepackage[justification=justified, format=plain]{caption}
\usepackage{subcaption}
\usepackage{physics}
\usepackage{mathtools}
\usepackage{amsmath,amsfonts,amssymb}
\usepackage[dvipsnames]{xcolor}
\usepackage{verbatim}
\usepackage{bbold}

\DeclareMathAlphabet\mathbfcal{OMS}{cmsy}{b}{n}

\usepackage[unicode=true,bookmarks=false,breaklinks=false,pdfborder={0 0 1},colorlinks=true]{hyperref}
\hypersetup{
 citecolor=blue,linkcolor=blue,urlcolor=blue}

 \newcommand{\overbar}[1]{\mkern 1.5mu\overline{\mkern-1.5mu#1\mkern-1.5mu}\mkern 1.5mu}

\begin{document}

\title{Position-Resolved Resonance Quantization for Lossy Cavities}
\author{Lucas Weitzel}
\email[present email: ]{lucas.weitzel@lkb.upmc.de}



\affiliation{Physikalisches Institut, Albert-Ludwigs-Universit\"at Freiburg, Hermann-Herder-Stra{\ss}e 3, D-79104 Freiburg, Germany}

\author{Andreas \surname{Buchleitner}}
\email[]{a.buchleitner@physik.uni-freiburg.de}

\author{Dominik \surname{Lentrodt}}
\email[]{dominik@lentrodt.com}

\affiliation{Physikalisches Institut, Albert-Ludwigs-Universit\"at Freiburg, Hermann-Herder-Stra{\ss}e 3, D-79104 Freiburg, Germany}
\affiliation{EUCOR Centre for Quantum Science and Quantum Computing, Albert-Ludwigs-Universit\"at Freiburg, Hermann-Herder-Stra{\ss}e 3, D-79104 Freiburg, Germany}

\date{\today}

\begin{abstract}

    Modern experiments in resonators are moving to ever more extreme quantum regimes, posing major challenges to established theoretical approaches, such as so-called few-mode models. While these models have driven major insights for traditional regimes, they are now hitting their limitations for highly open cavities and extended systems, as encountered in cavity experiments with molecules and solid-state systems. Here, we present a novel method that significantly extends the conceptual underpinning of these discrete-mode models, promoting them to a systematic treatment. We develop an ansatz which allows to quantize the resonator's resonances with position-resolved discrete modes, thus naturally incorporating losses in the formalism. Such a construction effectively unifies key ideas from pseudomodes and quantized quasi-normal modes theory. We further present a criterion for construction of the ansatz parameters at every point in space, and semi-analytically benchmark the resulting solution for a paradigmatic one-dimensional example resonator.

\end{abstract}

\maketitle


\section{Introduction}\label{sec:intro}


A cornerstone for the control of ever larger, engineered quantum objects --- from quantum computers \cite{Monroe1995} to quantum materials \cite{Jarc2023} --- is the confinement of the elementary constituents of light and matter. The conceptual underpinning of such endeavor is the idea of strong coupling. In the framework of quantum optics, strong light-matter coupling is achieved, for example, when a photon inside a resonator coherently interacts with a matter system, such as an atom. Many phenomena in this scenario can be described by the Jaynes-Cummings model \cite{Jaynes1963a,Larson2024}, where a single mode of the electromagnetic resonator field couples (near-)resonantly with a two-level system. The lossy version of this approach, here referred to as the open Jaynes-Cummings model, takes into account the leakage of photons from the single mode to an environment \cite{Agarwal2012, Puri1986}, thus describing more realistic resonators.

In the last decade, various experimental platforms in the field of cavity quantum electrodynamics (cQED) have emerged which depart from the simple setup encompassed by the Jaynes-Cummings model and/or of its multi-mode generalizations. For example, a parameter regime where the coupling strength is comparable to, or larger than the energy scales of the unperturbed matter system has been reached in many cQED experiments (see \cite{FriskKockum2019, Forn-Diaz2019} for reviews). Instead of atoms, also molecules, their chemical reactions \cite{Ebbesen2016,Thomas2019,GarciaVidal2021} and condensed-matter systems \cite{Appugliese2022,Jarc2023} are now being modified and controlled within electromagnetic cavities. Moreover, resonators featuring strong losses --- compared to the energy scale of a given resonant mode of interest --- such as sub-wavelength \cite{Ojambati2019, Benz2016}, plasmonic \cite{Hugall2018} and evanescent mode \cite{Vetsch2010} resonators, have been playing a significant role in cQED experiments, with potential applications thereof.

Regarding the theoretical description, one naturally has to accommodate the needs of these new regimes and setups. For instance, for sufficiently large coupling strengths, the well-known rotating-wave approximation is no longer applicable \cite{Shirley1965, Agarwal2012}. While the theory for the latter regime has been extensively discussed \cite{Braak2011, Agarwal2012, Nazir2018,DeLiberato2017}, the need for extensions has been pointed out especially for cavities with large losses, compared to their free spectral range \cite{Barnett1988a,Dutra2001,Khanbekyan2005,Hughes2018}. The central concern in this regard is the perturbative nature of the open Jaynes-Cummings model, as it describes the losses via master equations based upon the Markov approximation \cite{Breuer2002_BOOK,Carmichael1993, Cohen-Tannoudji1998b}.

Hence, multiple approaches have been suggested \cite{Garraway1997a, Garraway1997b, Lamprecht1999, Dutra2000, Dalton2001, Hackenbroich2002, Viviescas2003, Franke2019, Lentrodt2020, Medina2021, SanchezBarquilla2022, Lentrodt2023, Menczel2024} to tackle a central question: How can one extend the concept of discrete cavity modes to wider parameter regimes of cQED? This question is pertinent simply by the fact that, for a realistic cavity, the electromagnetic field features, \textit{a priori}, a continuum of modes. Consequently, the dynamics of quantum systems in those cases turns out to be, in general, a difficult problem to solve directly beyond perturbation theory \cite{Krimer2014, Strathearn2018, Breuer2002_BOOK}, and one often has to rely on numerical methods that discretize the mode-continuum \cite{Trivedi2021}. On the other hand, rigorous treatments featuring resonant modes offer a natural and efficient way to perform such a discretization, typically converging much faster than a direct sampling of the continuum \cite{Trivedi2021, Burkey1984, Rescigno1975, Maquet1983, Buchleitner1995, Reed1979iii, Menczel2024}, while still applicable to a broad range of light-matter coupling regimes \cite{Lambert2019,Medina2021,Lednev2024, Hornberger1998}.

Another important aspect in cQED is the spatial extension of a quantum emitter within a resonator. This aspect becomes ever more relevant especially in experiments involving the confinement and control of molecular \cite{Ebbesen2016} or solid-state \cite{Jarc2023} systems within cavities. Despite its importance, the position degree of freedom is often ignored in the aforementioned discretization models, which thus cannot \textit{continuously resolve} the position coordinate: Instead, only local couplings at isolated points in space are accounted for, and free model parameters are available to match the effective coupling \cite{Lambert2019,Medina2021,SanchezBarquilla2022,Lentrodt2023}. Methods which explicitly include the position-dependence, via first-principle techniques \cite{Viviescas2003, Viviescas2004, Franke2019}, rely on the assumption that the coupling between the resonator and the environment is weak, thus remaining perturbative in nature.

In our present contribution, we therefore set out to answer the question whether it is possible to construct an ``ideal'' extension of the open Jaynes-Cummings model, which features the following merits: First, that it faithfully represents a continuous, structured environment of a cavity by a possibly finite set of confined discrete modes. Second, that these confined modes are able to capture, according to a Lindblad master equation yet to be formulated, all non-Markovian effects of the dynamics of a quantum system coupled to the cavity, regardless of the light-matter or cavity-environment coupling strengths. Third, that such ``ideal'' parameters can be systematically derived, incorporating the spatial profile of a quantum emitter. We refer to the resulting effective mode structure as ``global (or generalized) pseudomodes'' (gPM) \cite{WeitzelThesis2025}, we apply the formalism to a generic cQED scenario, and spell out how to systematically solve the resulting mathematical problem.

The paper is structured as follows: In Sec.~\ref{sec:gpm_theory}, we specify the setup of interest. The essential idea is to reverse-engineer the free parameters of a discrete-mode ansatz, such as to guarantee a faithful modeling of the dynamics of a quantum system within a given cavity. In Sec.~\ref{sec:applications}, we apply and benchmark the method for a one dimensional example geometry, and compare it to alternative approaches. In the Appendices, we provide further details on the derivation of the method.

\section{Generalized pseudomode theory}
\label{sec:gpm_theory}

\subsection{Setup}
\label{sec:gpm_setup}

Let us begin by defining the cavity (or resonator): It consists of a region in space at whose boundaries the refractive index $n(\vb r)$ is discontinuous. Outside this region, $n(\vb r)$ is constant. We further assume that the cavity medium has no absorption or dispersion, meaning that $n(\vb r)$ is real and frequency-independent. We will stick to these assumptions here and throughout the paper.

Within the cavity, we assume a quantum system S which couples --- possibly strongly --- to the electromagnetic field. For simplicity and concreteness, the quantum system is assumed to be a set of $N$ emitters located at positions $\{\vb r_n\}$, ~$n=1,...\,, N$. We note, nevertheless, that our formalism does not rely on the specific structure of the matter degrees of freedom, but only on how they couple to the cavity field, as discussed in more detail in Sec.~\ref{sec:matching_condition}.

The cavity as defined above is, in general, leaky, meaning that confined photons can eventually escape to the outside. In such resonators, the electromagnetic field formally admits an expansion over a frequency-continuum of modes, which are defined at every point in space \cite{Glauber1991, Buhmann2012}. Throughout the present work, we will therefore refer to such a description of the field as the \textit{continuum representation} or \textit{continuum model} of the field.

Inspired by the so-called pseudomode formalism \cite{Garraway1997a,Garraway1997b,Dalton2001, Menczel2024}, we seek to replace the continuum representation of the electric field by an effective expansion into discrete modes, which is, nevertheless, able to reproduce the exact dynamics of the quantum system. We therefore employ the following ansatz for the electric field operator in the Heisenberg picture\footnote{We note that the Feshbach projection approach, employed in Refs.~\cite{Viviescas2003,Lentrodt2020}, is able to construct an analogous expression for the field expansion in Eq.~\eqref{eq:electric_field_FM}. The central difference in the former is that the corresponding modes are chosen explicitly as specific solutions of a wave equation. In our approach, on the other hand, they are functions yet to be determined, within the ansatz in Eq.~\eqref{eq:electric_field_FM}. The advantage of our approach, compared to the Feshbach projections, is that it does not restrict us to any pre-determined mode basis, but aims from the very start at optimizing it to its purpose describing the dynamics of the matter system by the Markovian master equation~\eqref{eq:master}.} \cite{WeitzelThesis2025},
\begin{align}
    \hat{\vb E}_\mathrm{gPM}(\vb r, t)= i\sum_\lambda\sqrt{\frac{\hbar}{2\epsilon_0}}\pmb\chi_\lambda(\vb r)\hat a_\lambda(t)+\text{h.c.}\label{eq:electric_field_FM}\,,
\end{align}
where $\{\pmb\chi_\lambda(\vb r)\}$ is a --- \textit{a priori} unspecified --- discrete set of confined modes. The ladder operators $\{\hat a_\lambda\}$ act on the subspace of the mode with index $\lambda$, and satisfy bosonic commutation relations,
\begin{align}
    \label{eq:canonical_commutation}
    [\hat a_\lambda, \hat a_{\lambda'}]&=[\hat a^\dagger_\lambda, \hat a^\dagger_{\lambda'}]=0,\;\;\forall\;\lambda,\lambda',\\
    \label{eq:canonical_commutation2}
    [\hat a_{\lambda}^{\strut}, \hat a^\dagger_{\lambda'}]&=\delta_{\lambda\lambda'}.
\end{align}

We refer to our proposed representation as \textit{generalized} or \textit{global} pseudomodes (gPM). The name stems from the fact that our approach departs from the concept of ``standard" pseudomodes in two aspects: First, the standard pseudomode operators are not directly associated with the electric field, as we do in Eq.~\eqref{eq:electric_field_FM}, but work solely as auxiliary modes. Second, our ansatz features a continuous parametrization of the modes $\pmb\chi_\lambda(\vb r)$ by the position coordinate.

Our goal with the gPM ansatz is to represent the time evolution of the entire cavity system --- that is, the quantum system S plus the gPMs --- via a Markovian master equation of the form \cite{Breuer2002_BOOK}
\begin{align}
    &\partial_t{\hat\rho}_\mathrm{S+gPM}(t)\nonumber\\&=-\frac i\hbar [\hat H_\mathrm{S+gPM}, \hat\rho_\mathrm{S+gPM}(t)]+\mathcal D_{\rm gPM}[\hat\rho_\mathrm{S+gPM}(t)],\label{eq:master}
\end{align}
where $\hat\rho_\mathrm{S+gPM}$ is the density matrix that represents the combined system. The free Hamiltonian in Eq.~\eqref{eq:master} is given by
\begin{align}
    \label{eq:free_Hamiltonian}
    \hat H_\mathrm{S+gPM} = \hat H_\mathrm{S} + \hat H_\mathrm{gPM} + \hat H_\mathrm{gPM}^\mathrm{int},
\end{align}
where we leave the quantum system's free Hamiltonian $\hat H_{\mathrm{S}}$ unspecified. The Hamiltonian $\hat H_\mathrm{gPM}$ in Eq.~\eqref{eq:free_Hamiltonian} is associated with the free gPM system, i.e., not interacting with S, and reads
\begin{align}
    \label{eq:hamiltonian_pseudomodes}
    \hat H_\mathrm{gPM}=\sum_{\lambda,\lambda'}\hbar\omega_{\lambda\lambda'}^{\strut}\hat a^\dagger_\lambda \hat a_{\lambda'}^{\strut},
\end{align}
where the parameters $\{\omega_{\lambda\lambda'}\}$ account both for the cavity mode frequencies (when $\lambda=\lambda'$) and for effective, excitation-conserving couplings between the cavity modes (when $\lambda\neq\lambda'$). Such a coupling can be thought of as encompassing environment-mediated excitation exchange between different modes. Furthermore, since $\hat H_{\rm gPM}$ is Hermitian, $\{\omega_{\lambda\lambda'}\}$ have to be the elements of a Hermitian matrix of parameters.

The interaction Hamiltonian $\hat H_\mathrm{gPM}^\mathrm{int}$ between the matter system S and the gPMs is set to be of the form
\begin{align}
    \label{eq:gpm_couplingH}
    \hat{H}_\mathrm{gPM}^\mathrm{int} = \sum_{n=1}^N \hat{\mathbf{E}}_{\rm gPM}(\mathbf{r}_n) \cdot \hat{\mathbfcal{O}}_{\mathrm{S},n}, \,
\end{align}
where $\{\hat{\mathbfcal{O}}_{\mathrm{S},n}\}$ are the vector operators associated, respectively, with the $n$-th emitter in the system S. For concreteness, $\hat{H}_\mathrm{gPM}^\mathrm{int}$ in Eq.~\eqref{eq:gpm_couplingH} is assumed to be in the dipolar form, such the system operators can be written as $\hat{\mathbfcal{O}}_{\mathrm{S},n}=\mathbf{d}_n \hat{\sigma}_n^+ + \;\mathrm{h.c.}$, where $\mathbf{d}_n$ is the $n$-th emitter's transition dipole moment.

Lastly, the superoperator $\mathcal D_{\rm gPM}$ in Eq.~\eqref{eq:master} accounts for losses or dissipation, mediated through the gPMs, from the cavity system to the environment, and has the form \cite{Breuer2002_BOOK}
\begin{align}
    \mathcal D_{\rm gPM}[\,\bullet\,]=&\sum_{\lambda,\lambda'}\kappa_{\lambda\lambda'}\left[\hat a^{\strut}_\lambda \bullet \hat a^\dagger_{\lambda'}-\frac12\{\hat a_\lambda^\dagger \hat a^{\strut}_{\lambda'},\bullet\}\right]\nonumber\\
    +&\sum_{\lambda,\lambda'}\gamma_{\lambda\lambda'}\left[\hat a^\dagger_\lambda \bullet\hat a^{\strut}_{\lambda'}-\frac12\{\hat a^{\strut}_\lambda \hat a^\dagger_{\lambda'}, \bullet\}\right],\label{eq:Lindbladian}
\end{align}
where $\{\kappa_{\lambda\lambda'}\}$ and $\{\gamma_{\lambda\lambda'}\}$ are parameters yet to be determined. We take $\mathcal D_{\rm gPM}$ not to be in diagonal form, to remain general, and, consequently, $\kappa_{\lambda\lambda'}$ and $\gamma_{\lambda\lambda'}$ are labeled by two indices. Their diagonal ($\lambda=\lambda'$) terms encode spontaneous emission and absorption of the gPMs, respectively, while the off-diagonal ($\lambda\neq\lambda'$) terms encompass cross-interaction processes, such as interferences between different decay channels. In order to preserve the positive semi-definiteness of $\hat\rho_\mathrm{S+gPM}$ throughout the system's time-evolution, we note that $\{\kappa_{\lambda\lambda'}\}$ and $\{\gamma_{\lambda\lambda'}\}$ are further required to be the elements of positive semi-definite matrices. We also observe that, with these assumptions, we impose the gPMs to have the interpretation of actual physical entities, a requirement which is not necessary in general: Pseudomode ansatzes featuring unphysical master equations and/or non-Hermitian analogs of the electric field operator in Eq.~\eqref{eq:electric_field_FM} have also been investigated in the literature \cite{Lamprecht1999, Lambert2019,Pleasance2020,Menczel2024}.

With the structure of the master equation given by Eq.~\eqref{eq:master}, we thus incorporate, in the discrete modes, an open system evolution which mimics the effect of the unitarily evolving continuum field on the dynamics of a matter system. In other words, this formulation can be regarded as a separation of the structured, continuous environment into a flat continuous part and the set of discrete cavity modes, which couple weakly to each other. The discrete modes, in turn, couple --- possibly strongly --- to the matter system.

To understand the gPMs, it is important to note that the dynamical equation Eq.~\eqref{eq:master} --- with both Hamiltonian and dissipation terms containing free parameters --- and the field expansion in Eq.~\eqref{eq:electric_field_FM} should be considered merely as tools do describe the dynamics of the quantum system S \cite{Tamascelli2018,Lambert2019,Medina2021}. That is, these modes form an effective basis set, yet to be determined, and are required to mimic, by construction, the effect of the continuous field onto the emitters. The Markovian character of the Lindblad master equation~\eqref{eq:master} is then \textit{not an approximation}, but a constructive element of the ansatz~\cite{Tamascelli2018,Menczel2024}. 

Eqs.~(\ref{eq:electric_field_FM} --~\ref{eq:Lindbladian}) thus specify our ansatz for the cavity system, with free parameters $\pmb\chi_{\lambda}(\vb r)$, $\omega_{\lambda\lambda'}$, $\kappa_{\lambda\lambda'}$ and $\gamma_{\lambda\lambda'}$. In order to provide a systematic way to recover the latter, we introduce, in the following subsection, an important criterion that guarantees that the mathematical structure considered above is indeed able to capture the dynamics of the quantum system within the cavity.

\subsection{Frequency domain matching condition}
\label{sec:matching_condition}

A key assumption behind our formalism --- and of pseudomode ansatzes in general --- is that the advanced and retarded two-point correlators of the environment's coupling operators be sufficient to determine the dynamics of the matter system in the resonator \cite{Tamascelli2018, Lambert2019, Menczel2024}. In Appendix~\ref{appendix:matching condition}, we provide a more precise formulation of such a requirement and also show that it can be equivalently formulated, in a simpler way compared to the previous literature, in the \textit{frequency domain}. We refer to the latter form of the equivalence statement as \textit{matching condition}, which we briefly summarize in the following, and we discuss how it can be exploited in the gPM formalism.

For the present discussion, it is sufficient to define only the advanced two-point correlator, which reads\footnote{The retarded correlator differs from the advanced one by having the first and second time arguments swapped.}
\begin{align}
    \mathbf{C}_{\mathrm{adv}}(\vb r, \vb r', \tau) \equiv i\langle\hat{\mathbf{E}}(\vb r, t+\tau) \otimes \hat{\mathbf{E}}(\vb r', t)\rangle \,,\label{eq::gpm_corr_time}
\end{align}
for $t,\tau\geq 0$, where $\otimes$ denotes the dyadic product between two vectors. The operator $\hat{\mathbf{E}}$ --- without any subscripts --- represents the electric field in either the continuous or the discrete representation. The expectation value is computed with respect to an \textit{decoupled} (from the matter degrees of freedom) initial state $\hat\rho_\mathrm{field}(0)$ of the field, that is $\expval{\;\bullet\;}\equiv\Tr[\;\bullet\;\hat\rho_\mathrm{field}(0)]$, where the field's state is taken to be stationary \cite{Menczel2024}. The latter attribute, in turn, is defined by the property of making the correlator in Eq.~\eqref{eq::gpm_corr_time} depend only on the time \textit{difference} $\tau$, and not on the particular instant $t$. For this reason, we fix $t=0$ in Eq.~\eqref{eq::gpm_corr_time} from now on. 

For our purposes, it will also be necessary to introduce the one-sided Fourier transform of the advanced two-point correlator. We define it as
\begin{align}
    \tilde{\mathbf{C}}_{\mathrm{adv}}(\vb r, \vb r', \omega)&\equiv \int_{0}^{\infty}\dd\tau \;\mathbf C_{\mathrm{adv}}(\vb r, \vb r', \tau)e^{i\omega \tau}.\label{eq:advanced_correlator_spectrum}
\end{align}
The one-sided Fourier transform is considered instead of the standard definition because the advanced correlator Eq.~\eqref{eq::gpm_corr_time} is defined for positive delays $\tau$. The sign convention for the exponential in Eq.~\eqref{eq:advanced_correlator_spectrum} is motivated in Appendix~\ref{appendix:matching condition}.

With the definitions in Eqs.~\eqref{eq::gpm_corr_time} and~\eqref{eq:advanced_correlator_spectrum}, we are now able to state the matching condition. It reads as follows \cite{Menczel2024}: If the imaginary part of the advanced frequency domain correlator defined in Eq.~\eqref{eq:advanced_correlator_spectrum} computed in the gPM approach matches the one computed for the actual continuous spectrum, that is, if
\begin{align}
    \Im\tilde{\vb C}^{\mathrm{gPM}}_{\mathrm{adv}}(\vb r, \vb r', \omega) \overset{!}{=} \Im\tilde{\vb C}^{\mathrm{cont}}_{\mathrm{adv}}(\vb r, \vb r', \omega) \,,\label{eq:matching_global_pseudomodes_to_green}
\end{align}
for all frequencies $\omega$, and at all positions $\vb r,\vb r'$ inside the resonator's volume, then the time evolution of the reduced (matter) system state $\hat\rho_\mathrm{S}(t)$ is the same in both representations. In other words, Eq.~\eqref{eq:matching_global_pseudomodes_to_green} is a sufficient and necessary condition \cite{Menczel2024} for the gPMs to completely capture the dynamics of a quantum system which, formally, interacts with the (continuous) cavity field.

The matching condition~\eqref{eq:matching_global_pseudomodes_to_green} nevertheless relies on two fundamental assumptions \cite{Menczel2024}:
\begin{itemize}
    \item[(i)] the stationary state of the environment --- whether the gPMs or the continuum field --- is Gaussian; and
    \item[(ii)] the coupling Hamiltonian [see, e.g., Eq.~\eqref{eq:gpm_couplingH}] is bilinear, in each representation, in both the matter system's and environment operators.
\end{itemize}

Regarding point (i), we mean, by Gaussian, states which are completely defined by the two-point correlators of the associated operators. Although such a property may seem quite a strong physical restriction to an environment's state at a first glance, it actually encompasses many states which are commonly dealt with in physical setups across the literature \cite{Breuer2002_BOOK, Agarwal2012}. As examples of Gaussian stationary states, one can mention the vacuum and thermal equilibrium states of the  electromagnetic field \cite{Breuer2002_BOOK}.

In the case of point (ii), the interaction Hamiltonian of the gPM ansatz [see Eq.~\eqref{eq:gpm_couplingH}] already has the required structure, while the coupling in the continuum representation was so far not specified. Hence, in order to comply with (ii), we assume that the quantum system S interacts with the electric field in the continuum model via dipolar coupling, with the corresponding interaction Hamiltonian having the same structure as Eq.~\eqref{eq:gpm_couplingH}. Nevertheless, we note that the required bilinearity is not restricted to the latter type of interaction, but is also realized, for example, in the case of magnetic dipole coupling \cite{Buhmann2012} or for coupling to the vector potential appearing in other gauge representations.

Let us now discuss the physical meaning of the matching condition~\eqref{eq:matching_global_pseudomodes_to_green}. The object on the right-hand side, evaluated for a continuum of modes, is directly related to the classical electric field's Green's dyadic. The latter is defined as the solution of \cite{Buhmann2012}
\begin{align}
    \left[\curl\curl-n^2(\vb r)\frac{\omega^2}{c^2}\right]\mathbf G(\vb r,\vb r',\omega)=\vb I\delta(\vb r-\vb r').\label{eq:equation_green}
\end{align}
where $\vb I$ is the identity matrix. For the vacuum state of the field, that is $\hat\rho_{\rm field}=\ketbra{0}$, one can show \cite{WeitzelThesis2025} that
\begin{align}
    \label{eq:correlator_to_green}
    \Im\tilde{\vb C}^{\rm cont}_{\mathrm{adv}}(\vb r, \vb r', \omega)&=\frac{\hbar}{\epsilon_0 c^2}\theta(\omega){\omega}^2\Im\vb G(\vb r,\vb r',\omega) \,,
\end{align}
where $\theta(\omega)$ is the Heaviside function.

The relation in Eq.~\eqref{eq:correlator_to_green} means that $\Im\tilde{\vb C}^{\rm cont}_{\mathrm{adv}}(\vb r, \vb r', \omega)$ encodes the information about the light-matter interaction at position $\vb r$ originating from the field whose source is located at position $\vb r'$. In particular, when evaluated at coincidence $\vb r=\vb r'$, Eq.~\eqref{eq:correlator_to_green} is proportional to the spectral density of the cavity field. With such an interpretation, the matching condition can also be understood as matching the spectral densities of the continuous and discrete environments \cite{Medina2021,Lambert2019}, in order for the gPMs to yield the same dynamics of the matter system as in the continuum framework.

Regarding the gPM-analog of Eq.~\eqref{eq:correlator_to_green}, which appears on the left-hand side of Eq.~\eqref{eq:matching_global_pseudomodes_to_green}, it can be computed analytically by plugging into Eq.~\eqref{eq:advanced_correlator_spectrum} the gPM expansion for the electric field [Eq.~\eqref{eq:electric_field_FM}]. This calculation is performed in Appendix~\ref{appendix:correlator_pseudomodes}, where we show that
\begin{align}
    &\Im\tilde {\vb C}^\mathrm{gPM}_{\mathrm{adv}}(\vb r, \vb r', \omega)\nonumber\\&=\frac{\hbar}{2\epsilon_0}\Im\left[\sum_{\lambda,\lambda'}\boldsymbol\chi_{\lambda}(\vb r)(\mathbb H-\omega)^{-1}_{\lambda\lambda'}\boldsymbol\chi^\dagger_{\lambda'}(\vb r')\right] \,, \label{eq:correlator_global_pseudomode}
\end{align}
which we refer to as a \textit{gPM expansion} of the spectral density, due to the fact that the modes $\boldsymbol\chi_{\lambda}(\vb r)$ from Eq.~\eqref{eq:electric_field_FM} appear explicitly. In Eq.~\eqref{eq:correlator_global_pseudomode}, the dagger denotes the complex transpose of a \textit{vector}, and $\mathbb H_{\lambda\lambda'}\equiv \mathbb \omega_{\lambda\lambda'}- i(\kappa_{\lambda\lambda'}-\gamma_{\lambda\lambda'})/2$ is a matrix of the free parameters appearing in the gPM Hamiltonian~\eqref{eq:hamiltonian_pseudomodes} and in the master equation~\eqref{eq:master}. Such a matrix can also be interpreted as a representation --- in the basis of states associated with the gPM ladder operators $\{\hat a_\lambda\}$ [see Eq.~\eqref{eq:electric_field_FM}] --- of an effective non-Hermitian Hamiltonian \cite{Maquet1983, Buchleitner1995, Reed1979iii, Moiseyev2011, Rotter2009, El-Ganainy2018} that dictates the evolution of the free gPMs.

Eq.~\eqref{eq:matching_global_pseudomodes_to_green} thus provides us with a powerful tool, since it allows us to systematically recover the parameters $\boldsymbol\chi_{\lambda}(\vb r)$ and $\mathbb H$, as described in the next section. 

\subsection{From pseudomodes to pole expansions (and back)}
\label{sec:from_pseudomodes_to_pole}

In the following, we reverse engineer the parameters $\boldsymbol\chi_{\lambda}(\vb r)$ and $\mathbb H$ by exploiting the matching condition~\eqref{eq:matching_global_pseudomodes_to_green}. Essentially, the approach can be summarized in three steps: In Sec.~\ref{sec:gPM_Hermitization_cond}, we first manipulate Eq.~\eqref{eq:correlator_global_pseudomode} to cast it as a pole expansion. Then, in Sec.~\ref{sec::gPM_poleExp}, we connect the latter form to resonant mode expansions from classical resonance theory, by evoking the matching condition~\eqref{eq:matching_global_pseudomodes_to_green}. Finally, in Sec.~\ref{sec:gpm_to_qnm}, this connection leads us to a single matrix equation, that needs to be solved under certain constraints.

\subsubsection{Hermitization condition}\label{sec:gPM_Hermitization_cond}

We first follow \cite{Lentrodt2023} to write Eq.~\eqref{eq:correlator_global_pseudomode} in a more convenient form by diagonalizing the matrix $\mathbb H$. 
Such a diagonalization can be interpreted as a change of basis from the gPMs to states of the field known as resonant \cite{Reed1979iii, Kukulin1989} or quasi-normal modes \cite{Lalanne2018,Kristensen2020}, which have been extensively studied in the literature. Another related concept are the constant-flux states from laser theory \cite{Tureci2006}.

Hence, we proceed by defining an invertible matrix $V$ such that $(V\mathbb H V^{-1})_{\mu\nu}=\tilde\Omega_{\mu}\delta_{\mu\nu}$. The diagonalization of $\mathbb H$ in Eq.~\eqref{eq:correlator_global_pseudomode} yields
\begin{align}
    \Im\tilde{\vb C}^\mathrm{gPM}_{\mathrm{adv}}(\vb r,\vb r',\omega)
    &=\frac{\hbar}{2\epsilon_0}\Im\left[\sum_\mu\overbar{\vb g}^*_{\mu}(\vb r)(\tilde\Omega_\mu-\omega)^{-1}\tilde{\vb g}^{\mathrm{T}}_{\mu}(\vb r')\right]\,,\label{eq:diag_propagator}
\end{align}
where the superscript T denotes the transpose of a \textit{vector}, and we have introduced the functions
\begin{align}
    \overbar{\vb g}_\mu (\vb r) &\equiv \sum_{\lambda} (V^\dagger)^{-1}_{\mu\lambda}\pmb \chi^*_{\lambda}(\vb r),\label{eq:g_mode1}
    \\
    \tilde{\vb g}_{\mu}(\vb r) &\equiv \sum_{\lambda}V^{\strut}_{\mu\lambda}\pmb\chi^*_{\lambda}(\vb r) \,.\label{eq:g_mode2}
\end{align}

Eq.~\eqref{eq:diag_propagator} thus has the form of (the imaginary part of) a pole expansion, whose poles are the --- generally complex --- eigenvalues $\tilde\Omega_\mu$ of the matrix $\mathbb H$. These eigenvalues, in turn, can be interpreted as the characteristic frequencies of the modes we denote by $\overbar{\vb g}_\mu (\vb r)$ and $\tilde{\vb g}_{\mu}(\vb r)$. Being associated with a complex frequency means that, besides having a natural oscillation, the latter modes will have a decay rate. The precise relationship between $\tilde\Omega_\mu$ and $\tilde{\vb g}_\mu(\vb r)$ and $\overbar{\vb g}_\mu(\vb r)$ is resolved in Secs.~\ref{sec::gPM_poleExp} and~\ref{sec:gpm_to_qnm} below, in particular by Eqs.~(\ref{eq:wave_equation_qnm},~\ref{eq:poles_id},~\ref{eq:modes_id}).

As mentioned previously in Sec.~\ref{sec:gpm_setup}, the problem with the gPM expansion in Eq.~\eqref{eq:correlator_global_pseudomode} is that both the matrix $\mathbb H$ and the modes $\pmb\chi_{\lambda}(\vb r)$ are still unknown. Hence, it is unclear, at this level of the discussion, how to construct the transformation matrix $V$. Instead, our approach works in reverse, that is, we construct the gPM parameters by knowing $\overbar{\vb g}_\mu (\vb r)$, $\tilde{\vb g}_{\mu}(\vb r)$, $\tilde\Omega_\mu$, as we detail throughout this subsection. This task has previously been identified as an open problem in the standard pseudomodes framework \cite{Garraway1997a,Garraway1997b} and was termed ``undiagonalization'' in \cite{SanchezBarquilla2022}.

Continuously fulfilling the matching condition~\eqref{eq:matching_global_pseudomodes_to_green} along the position coordinate, as demanded for the gPMs $\pmb\chi_\mu(\vb r)$, places a strong constraint on the transformation matrix $V$ and, therefore, on the gPMs' parameters. In particular, a constraint on $V$ can be derived by combining Eqs.~\eqref{eq:g_mode1} and \eqref{eq:g_mode2} into \cite{WeitzelThesis2025}
\begin{align}
    \tilde{\vb g}_\mu(\vb r)=\sum_{\lambda\nu}V^{\strut}_{\mu\lambda}V_{\lambda\nu}^\dagger\overbar{\vb g}_\nu(\vb r) \,.\label{eq:matrix_squaree_root}
\end{align}
We will refer to this expression as \textit{Hermitization condition}, as a central constraint defining the ``undiagonalization'' transformation.

To gain additional insight into the origin of this relation, we compare the two expansions, Eqs.~\eqref{eq:correlator_global_pseudomode} and \eqref{eq:diag_propagator}. In the latter, the poles can also be understood as the elements of a diagonal matrix $\text{diag}(\{\tilde\Omega_\mu\})$. Eq.~\eqref{eq:correlator_global_pseudomode}, on the other hand, has the crucial feature of having the modes $\{\boldsymbol{\chi}_\lambda(\vb r)\}$ appearing together with their \textit{complex transpose}. This property and, by extension, the Hermitization condition, are a direct consequence of the Hermitian nature of the gPM ansatz from Eq.~\eqref{eq:electric_field_FM}. We note that there are schemes where the approach featuring both $\pmb\chi_\mu(\vb r)$ and $\pmb\chi_\mu^\dagger(\vb r)$ can be circumvented, by non-Hermitian expansions of the electric field \cite{Lamprecht1999, Pleasance2020,  Lambert2019, Menczel2024} that can correctly describe the dynamics of quantum systems. The trade-off is that the local environment coupled to the emitters is represented by unphysical modes. In our approach, the requirement of a physically meaningful, i.e., Hermitian, mode expansion of the electric field operator leads to Eq.~\eqref{eq:matrix_squaree_root}.

The Hermitization condition is beneficial as it can be used to find the transformation matrix $V$. However, it is only useful if the parameters of the pole expansion --- that is, the functions $\overbar{\vb g}_\mu (\vb r)$, $\tilde{\vb g}_{\mu}(\vb r)$ and poles $\tilde\Omega_\mu$ --- can be calculated  independently. As we show in the following, the latter is indeed the case, due to the relation between the correlator and the Green's function given by Eq.~\eqref{eq:correlator_to_green}.

\subsubsection{Pole expansion in the continuum theory}\label{sec::gPM_poleExp}

The modes $\overbar{\vb g}_\mu (\vb r)$, $\tilde{\vb g}_{\mu}(\vb r)$ and poles $\tilde\Omega_\mu$ appearing in Eq.~\eqref{eq:diag_propagator} can be extracted from the object $\Im\tilde{\vb C}^{\mathrm{cont}}_{\mathrm{adv}}(\vb r, \vb r', \omega)$, appearing on the right-hand side of the matching condition Eq.~\eqref{eq:matching_global_pseudomodes_to_green}. We establish such a correspondence by also casting the latter as a pole expansion \cite{Lambropoulos2000}, but, this time, in terms of objects that are directly obtained from Maxwell's equations. However, as we discuss in the following, such expansions can never be exact for the vacuum case, and, thus, we will need to resort to certain approximations.

Let us focus on the Green's function appearing on the right-hand side of Eq.~\eqref{eq:matching_global_pseudomodes_to_green}. It is known from the classical resonance theory of electromagnetic cavities \cite{Kristensen2020,Lalanne2018} (see Appendix~\ref{appendix:Green_pole_expansion}) that\footnote{We note that in the pole expansion Eq.~\eqref{eq:green_qnm2}, the quasi-normal modes appear twice in the residues \textit{without} a complex conjugation \cite{Doost2014}. This important feature --- as already noted in \cite{Garraway1997a,Lentrodt2023} --- implies that the residues can be complex, which evidences the necessity of the Hermitization condition~\eqref{eq:matrix_squaree_root}, in order to comply with the properties of the electric field operator.}
\begin{align}
    \omega^2\Im\vb G(\vb r, \vb r',\omega)=\frac{c^2}{2}\Im\left[\sum_{\mu=-\infty}^{+\infty}\frac{\tilde\omega_\mu\tilde{\vb f}_\mu(\vb r)\otimes\tilde{\vb f}_\mu(\vb r')}{\tilde\omega_\mu-\omega}\right], \label{eq:green_qnm2}
\end{align}
where $\tilde{\vb f}_\mu(\vb r)$ are the resonant states or \textit{quasi-normal modes} (QNMs) \cite{Ching1998}. They are defined as solutions of the wave equation \cite{Kristensen2020, Sauvan2022}
\begin{align}
    \curl\curl\tilde{\vb f}_\mu(\vb r)-n^2(\vb r)\frac{\tilde\omega_\mu^2}{c^2}\tilde{\vb f}_\mu=0,\label{eq:wave_equation_qnm}
\end{align}
subjected to purely outgoing-wave boundary conditions \cite{Kristensen2020, Sauvan2022, Taylor2012_BOOK, Ho1983method, Siegert1939}. These boundary conditions at $\vb r\to\infty$ --- also known as Silver-Müller radiation condition \cite{Kristensen2020} --- are expressed mathematically as
\begin{align}
    \lim_{|\vb r|\to\infty}\hat{\vb r}\times\tilde{\vb f}_\mu(\vb r)= -in_{\mathrm{B}}\frac{\tilde\omega_\mu}c\tilde{\vb f}_\mu(\vb r).\label{eq:silver_mueller}
\end{align}
One consequence of this boundary condition is that the eigenfrequencies $\tilde\omega_\mu$ are complex in general \cite{Ching1998, Kristensen2020}. Mathematically, such a property is a consequence of the QNMs being solutions of a non-Hermitian problem, implying usually distinct properties compared to the Hermitian case. For example, the set of QNMs constitutes an overcomplete basis, meaning that the QNMs are linearly \textit{dependent} on each other and, nevertheless, satisfy a completeness relation. This and other relevant properties are discussed in more detail in Appendix~\ref{appendix:qnm_properties}, and they are important tools to help us solve the Hermitization condition Eq.~\eqref{eq:matrix_squaree_root}, as we show in Sec.~\ref{sec:gpm_to_qnm}. 

Plugging the expansion of Eq.~\eqref{eq:green_qnm2}, for the imaginary part of the Green's function, into Eq.~\eqref{eq:correlator_to_green}, one naturally arrives at
\begin{align}
    \Im\tilde{\vb C}^\mathrm{cont}_{\mathrm{adv}}&(\vb r, \vb r', \omega)\nonumber\\=&\frac{\hbar}{2\epsilon_0}\theta(\omega)\text{Im}\left[\sum_{\mu=-\infty}^{+\infty}\frac{\tilde\omega_\mu\tilde{\vb f}_\mu(\vb r)\otimes\tilde{\vb f}_\mu(\vb r')}{\tilde\omega_\mu-\omega}\right].\label{eq:non-meromorphic}
\end{align}

By comparing Eqs.~\eqref{eq:diag_propagator} and~\eqref{eq:meromorphic_approx}, we observe that the matching condition cannot be exactly satisfied. The reason for such incompatibility is that the Heaviside function appearing Eq.~\eqref{eq:non-meromorphic} makes $\Im\tilde{\vb C}^\mathrm{cont}_{\mathrm{adv}}(\vb r, \vb r', \omega)$ non-smooth\footnote{More generally, the object in Eq.~\eqref{eq:non-meromorphic} is non-meromorphic, and it is known that non-meromorphic complex functions do not admit pole expansions, by the Mittag-Leffler theorem \cite{Ablowitz2003}.} in the vicinity of $\omega=0$. Consequently, the latter object does not admit an exact pole expansion, that is, the form of Eq.~\eqref{eq:diag_propagator}. Physically, this restriction means that our gPM ansatz [see Sec.~\ref{sec:gpm_setup}] cannot exactly mimic the coupling to a continuous electric field, which was already observed in \cite{Tamascelli2018}.

To circumvent such a hindrance, we instead rely on constructing an \textit{approximate} pole expansion, which we refer to as a \textit{meromorphic approximation}. We choose to approximate Eq.~\eqref{eq:non-meromorphic} as
\begin{align}
    \text{Im}\tilde{\vb C}^\mathrm{cont}_{\mathrm{adv}}(\vb r, \vb r', \omega)\approx\frac{\hbar}{2\epsilon_0}\text{Im}\left[\sum_{\mu>0}^{+\infty}\frac{\tilde\omega_\mu\tilde{\vb f}_\mu(\vb r)\otimes\tilde{\vb f}_\mu(\vb r')}{\tilde\omega_\mu-\omega}\right],\label{eq:meromorphic_approx}
\end{align}
where the truncation to $\mu>0$ indicates that we are only summing over the poles with positive real part [$\mathrm{Re}(\tilde\omega_\mu)>0$]. The physical implications of this truncation, and how it approximates the low-frequency behavior of the correlator, will be discussed further in Sec.~\ref{sec:applications}. Essentially, the meromorphic approximation has the role of smoothing the function in the vicinity of $\omega=0$. We reinforce, though, that the approximation by truncating the summation Eq.~\eqref{eq:non-meromorphic} is a \textit{choice}, and other meromorphic approximations can be achieved via different constructions, e.g., by replacing the step function $\theta(\omega)$ by a smooth alternative, so that a truncation of the sum is not needed.\footnote{In fact, the need of a meromorphic approximation, as in Eq.~\eqref{eq:meromorphic_approx}, is a consequence of assuming the initial state of the field to be the vacuum, and does not necessarily persist in other cases. For example, one can show that $ \text{Im}\tilde{\vb C}^\mathrm{cont}_{\mathrm{adv}}(\vb r, \vb r', \omega)$ is indeed meromorphic if the initial field state fulfills thermal equilibrium statistics~\cite{WeitzelThesis2025}.}

We can now finally exploit the matching condition by identifying Eq.~\eqref{eq:meromorphic_approx} with Eq.~\eqref{eq:diag_propagator}. In the following, we will exploit this correspondence to recover $\overbar{\vb g}_\mu (\vb r)$, $ \tilde{\vb g}_{\mu}(\vb r)$ and $\tilde\Omega_\mu$, which so far were left free in Eq.~\eqref{eq:diag_propagator}, and to solve the Hermitization condition Eq.~\eqref{eq:matrix_squaree_root} for $V$.

\subsubsection{Linking pseudomodes and quasi-normal modes}
\label{sec:gpm_to_qnm}

By the matching condition Eq.~\eqref{eq:matrix_squaree_root}, we demand Eqs.~\eqref{eq:diag_propagator} and~\eqref{eq:meromorphic_approx} to be the same. In order to fulfill this requirement, we demand a one-to-one correspondence between the poles of Eqs.~\eqref{eq:diag_propagator} and of~\eqref{eq:meromorphic_approx}, and, similarly, between their corresponding residues. In the case of poles, we naturally have
\begin{align}
    \label{eq:poles_id}
    \tilde\Omega_\mu = \tilde\omega_\mu,
\end{align}
so that we associate the complex eigenvalues $\tilde\Omega_\mu$ of the effective Hamiltonian $\mathbb H$, appearing in Eq.~\eqref{eq:diag_propagator}, with the QNM's eigenfrequencies $\tilde\omega_\mu$. 

Setting the residues in Eqs.~\eqref{eq:diag_propagator} and \eqref{eq:meromorphic_approx} to be the same leads to a relation between $\overbar{\vb g}_{\mu}(\vb r)$ and $\tilde{\vb g}_{\mu}(\vb r)$ and the QNMs $\tilde{\vb f}_{\mu}(\vb r)$. In the following, we opt to identify
\begin{align}
    \label{eq:modes_id}
    \tilde{\vb g}_\mu(\vb r)=\tilde\omega_\mu\tilde{\vb f}_\mu(\vb r),\;\;\;\overbar{\vb g}_\mu(\vb r)=\tilde{\vb f}^*_\mu(\vb r).
\end{align}
We note that, with Eq.~\eqref{eq:modes_id}, we made a choice of how to split the constant factor $\tilde\omega_\mu$ between the two modes. This choice is motivated purely by mathematical reasons, as it simplifies the calculations to be performed ahead, and has no physical implications.

We are now equipped with all the necessary ingredients to solve the Hermitization condition Eq.~\eqref{eq:matrix_squaree_root} for $V$. We again emphasize the crucial feature that the residues of the pole expansion in Eq.~\eqref{eq:meromorphic_approx} can be complex, which leads to a non-trivial constraint for the gPMs, demanded by the Hermitization condition. Hence, substituting Eq.~\eqref{eq:modes_id} into Eq.~\eqref{eq:matrix_squaree_root}, we have \cite{WeitzelThesis2025}
\begin{align}
    \tilde\omega_\mu\tilde{\vb f}_{\mu}(\vb r)=\sum_{\nu>0}T_{\mu\nu}\tilde{\vb f}^*_{\nu}(\vb r),\label{eq:hermiticity1}
\end{align}
with
\begin{align}
    T_{\mu\nu}\equiv\sum_{\lambda}V_{\mu\lambda}V_{\lambda\nu}^\dagger,\label{eq:hermiticity2}
\end{align}
where the restricted index domain of $\nu>0$ in Eq.~\eqref{eq:hermiticity1} is a consequence of the truncation of the series in Eq.~\eqref{eq:meromorphic_approx} to $\mu>0$. 

With Eqs.~\eqref{eq:hermiticity1} and~\eqref{eq:hermiticity2}, we promoted the Hermitization condition~\eqref{eq:matrix_squaree_root} to an explicit constraint on $V$, since the QNMs $\tilde{\vb f}_\mu(\vb r)$ can, in principle, be obtained by solving their corresponding wave equation [see Sec.~\ref{sec::gPM_poleExp}], or by other methods, for which a large toolbox is available \cite{Betz2021,Binkowski2020,Lalanne2018,Kristensen2020}.

We observe that the identification of modes as done in Eq.~\eqref{eq:modes_id} subsequently equips $T$ with the dimension of frequency, and $V$ with the dimension of square root of frequency. Nevertheless, this does not have any physical implications, as it does not affect the dimensionality of $\mathbb H$ and, consequently, of the gPM's free parameters, being simply a matter of convenience for the calculations performed in the next subsections.

\subsection{Approximate solution of the Hermitization condition}
\label{sec::gPM_approxSolHerm}

At this point, it is important to summarize what we have achieved so far. We began with the gPM ansatz presented in Sec~\ref{sec:gpm_setup}, whose field expansion [see Eq.~\eqref{eq:electric_field_FM}] and associated master equation [see Eq.~\eqref{eq:master}] contained free parameters to be determined. Then, we concluded that such a gPM ansatz associates a pole expansion, in the frequency domain, to the advanced correlator [see Eq.~\eqref{eq:diag_propagator}]. The pole expansion must fulfill an additional constraint, the Hermitization condition~\eqref{eq:matrix_squaree_root}. Finally, imposing the matching condition~\eqref{eq:matching_global_pseudomodes_to_green} to hold, we were able, with the approximation~\eqref{eq:meromorphic_approx}, to convert the Hermitization condition into Eqs.~\eqref{eq:hermiticity1} and~\eqref{eq:hermiticity2}, and to reduce our whole problem of finding the free parameters to solve these equations for $V$, given that the QNMs, introduced in Eq.~\eqref{eq:green_qnm2}, are known \textit{a priori}.

We dedicate this section to solve Eq.~\eqref{eq:hermiticity1} for the matrix $T$, with Eq.~\eqref{eq:hermiticity2} imposing additional constraints on the solution. $T$ is constructed explicitly in terms of the QNMs, and it subsequently allows us to recover all the free parameters in the gPM setup.

\subsubsection{Extended index domain solution}

First of all, we note that Eq.~\eqref{eq:hermiticity1} can be seen as expressing $\tilde\omega_\mu\tilde{\vb f}_\mu(\vb r)$ as a linear combination of other QNMs. As the QNMs form an \textit{overcomplete} basis, it suggests that Eq.~\eqref{eq:hermiticity1} can be satisfied, with multiple possible solutions for $T$. However, the linear dependence relation [see Eq.~\eqref{eq:overcompleteness} in the Appendix] involves the whole QNM set, that is for all $\mu\in\mathbb Z$, whereas Eq.~\eqref{eq:hermiticity1} encompasses the restricted index domain $\nu>0$. Consequently, only approximate solutions can be found. It has been conjectured \cite{Hoenders1978}, nevertheless, that the set of QNMs with $\mu>0$ is (over)complete on its own. If this is the case, then Eq.~\eqref{eq:hermiticity1} can be solved exactly.

In order to construct an approximate solution for $T$ from Eq.~\eqref{eq:hermiticity1}, we slightly modify our problem. We consider the same equation but with the index domain extended to all integers, such that all QNMs are included in the sum, that is
\begin{align}
    \tilde\omega_\mu\tilde{\vb f}_{\mu}(\vb r)=\sum_{\nu = -\infty}^{\infty} \tilde{T}_{\mu\nu}\tilde{\vb f}^*_{\nu}(\vb r).\label{eq:hermiticity_extended}
\end{align}
Eq.~\eqref{eq:hermiticity_extended} can then be solved exactly for $\tilde{T}$. As we detail in Appendix~\ref{appendix:qnm_properties}, the properties of the QNMs allow for two independent solutions \cite{WeitzelThesis2025},
\begin{align}
    \tilde{T}^{(1)}_{\mu\nu}=\tilde\omega_{\mu}\delta_{\mu,-\nu},\label{eq:T1}
\end{align}
and
\begin{align}
    \tilde{T}^{(2)}_{\mu\nu}=\frac{\tilde\omega_\mu}2\int_{\mathcal R}\dd^3\vb r\;n^2(\vb r)\tilde{\vb f}_\mu(\vb r)\cdot\tilde{\vb f}^*_\nu(\vb r),\label{eq:T2}
\end{align}
where $\mathcal R$ denotes the spatial volume of the resonator. We observe that $\tilde{T}^{(1)}$ is completely anti-Hermitian, that is, $\tilde{T}^\dagger=-\tilde{T}$, and that $\tilde{T}^{(2)}$ holds only for dispersionless and absorptionless media, where $n(\vb r)\in\mathbb R$.

We can thus create a whole family of solutions of Eq.~\eqref{eq:hermiticity_extended} from linear combinations of Eq.~\eqref{eq:T1} and Eq.~\eqref{eq:T2}, which reads\footnote{We highlight that Eq.~\eqref{eq:family} does not necessarily encompasses \textit{all} solutions of Eq.~\eqref{eq:hermiticity_extended}. Nevertheless, the accessible solutions already reveal a good performance in the application to a concrete case, as we will see in Sec.~\ref{sec:applications}.}
\begin{align}
    \tilde{T}_{\mu\nu}=(1-a)\tilde{T}^{(1)}_{\mu\nu}+a\tilde{T}^{(2)}_{\mu\nu},\;\;a\in\mathbb C.\label{eq:family}
\end{align}

The solutions in Eq.~\eqref{eq:family} will subsequently serve as the approximate solutions for Eq.~\eqref{eq:hermiticity1}, over the truncated index domain ($\mu,\nu>0$). Nevertheless, we still have to ensure that the constraint of Eq.~\eqref{eq:hermiticity2} is fulfilled. We discuss this aspect in the following subsection.

\subsubsection{From $\tilde{T}$ to $V$} 
To fully solve the Hermitization condition, one has to construct $V$ from $T$. For a solution of $V V^\dagger=T$ to exist, $T$ must be Hermitian and, in particular, positive semi-definite, as known from linear algebra \cite{Horn2012}. However, none of the matrices within the family of solutions for $\tilde{T}$ in Eq.~\eqref{eq:family} are even Hermitian, due to the presence of a $\tilde\omega_\mu$ factor in Eq.~\eqref{eq:T2}.

Hence, we have to proceed by finding a Hermitian matrix that well approximates one of the solutions encompassed by Eq.~\eqref{eq:family}. As we have the freedom to choose any complex $a$ in the latter equation, we take the matrix whose anti-Hermitian part is minimized.\footnote{By minimizing, we mean to choose $a$ such that all anti-Hermitian terms of the form of $\tilde T_{\mu\nu}^{(1)}$ in Eq.~\eqref{eq:family} are eliminated. Although $a=1$ seems like a straightforward choice, this is not always appropriate, as discussed further in Sec.~\ref{sec:applications} and in Appendix~\ref{appendix::analytical_solution_T}.} With the choice of a suitable $a$, we then approximate the resulting matrix as $\tilde{T} \approx \tilde{T}^{\text{H}}$, that is, we take the Hermitian part of the matrix as an approximation for itself.

Nevertheless, we note that only having a Hermitian solution does not guarantee yet the matrix to be positive semi-definite. Still, a general proof that the solutions obtained through Eq.~\eqref{eq:family} are indeed positive-definite remains to be found. We, therefore, leave the verification of this requirement to Sec.~\ref{sec:applications}, where we turn to a specific example.

Provided that $\tilde{T}^{\text{H}}$ is indeed positive-definite, it is then an approximate solution of Eq.~\eqref{eq:hermiticity1}, which can be systematically constructed from the QNMs, and that fulfills the constraint of Eq.~\eqref{eq:hermiticity2}. Using the latter, one can  construct, from $\tilde{T}^{\text{H}}$, the matrix $V$, which, in turn, allows one to recover the gPM parameters, $\pmb\chi_\mu(\vb r)$ and $\mathbb H$, from the QNMs $\tilde{\vb f}_\mu(\vb r)$ and its frequencies $\tilde\omega_\mu$. While an explicit construction of $V$ in terms of the QNMs is yet to be determined, there are several standard numerical techniques to compute matrix square roots \cite{Horn2012}.

\subsection{Recovering the gPM parameters}
\label{sec:recovering}

For the sake of clarity, we end this section by summarizing the approximations performed to satisfy the matching condition Eq.~\eqref{eq:matching_global_pseudomodes_to_green} with the gPM ansatz, and how the free parameters of the latter setup can be recovered with the solution of the Hermitization conditions, Eqs.~\eqref{eq:hermiticity1} and~\eqref{eq:hermiticity2}.

The starting point to recover the free parameters in the gPM ansatz is to the determine the QNMs $\tilde{\vb f}_{\mu}(\vb r)$ of a considered cavity geometry. They are obtained by solving the wave equation~\eqref{eq:wave_equation_qnm}, subjected to the Silver-Müller condition~\eqref{eq:silver_mueller}.

With the QNMs, we construct, via Eq.~\eqref{eq:family}, a family of solutions for $\tilde T$ that exactly satisfies Eq.~\eqref{eq:hermiticity_extended} for the index domain $\mu, \nu\in\mathbb Z$. We then truncate the domain to the indices $\mu,\nu>0$ and take the Hermitian part $\tilde T^{\rm H}$ of the resulting matrix, in order to comply with Eq.~\eqref{eq:hermiticity2}. Finally, we use $\tilde T^{\rm H}$ to write the imaginary part of the half-range Fourier transform of the advanced correlator as
\begin{align}
    \Im\tilde{\vb C}^{\rm gPM}_{\mathrm{adv}}(\vb r, \vb r',\omega)& = \frac{\hbar}{2\epsilon_0}\Im\left[\sum_{\mu>0,\nu>0}^\infty\frac{\tilde{\vb f}_{\mu}(\vb r)\otimes \tilde T^{\text{H}}_{\mu\nu}\tilde{\vb f}^*_{\nu}(\vb r')}{\tilde\omega_\mu-\omega}\right]\nonumber\\
    &\approx\Im\tilde{\vb C}^{\rm cont}_{\mathrm{adv}}(\vb r, \vb r',\omega).\label{eq:hermitized_pole_correlator}
\end{align}

The importance of the above expansion is two-fold: It is designed to approximate $\Im\tilde{\vb C}^{\rm cont}_{\mathrm{adv}}(\vb r, \vb r',\omega)$, found with the continuum approach, while, at the same time, being compatible with the structure demanded by the gPM ansatz, that is, having the form of (the imaginary part of) a pole expansion and satisfying the Hermitization condition Eq.~\eqref{eq:matrix_squaree_root}. The first property ensures that the matching condition Eq.~\eqref{eq:matching_global_pseudomodes_to_green} is fulfilled, at least approximately, while the second property allows us to compute the gPM free parameters, as we will discuss below. The quality of the aforementioned approximations will be investigated in more detail in Sec.~\ref{sec:applications}, where we focus on a concrete example.

We proceed by decomposing the approximate solution of Eq.~\eqref{eq:hermiticity1} as $\tilde T^{\rm H} = VV^\dagger$. The gPMs are then constructed from the QNMs either by inverting Eq.~\eqref{eq:g_mode1}, which gives
\begin{align}
    {\pmb \chi}^*_{\lambda}(\vb r)&=\sum_{\nu}V^\dagger_{\lambda\nu}\tilde{\vb f}^*_\nu(\vb r),\label{eq:chi_mode1}
\end{align}
or, similarly, by inverting Eq.~\eqref{eq:g_mode2}, which yields
\begin{align}
    {\pmb\chi}^*_{\lambda}(\vb r) &=\sum_{\nu}(V^{-1})_{\lambda\nu}\tilde\omega_\nu\tilde{\vb f}_\nu(\vb r) \,.\label{eq:chi_mode2}
\end{align}
These expressions are equivalent constructions if the transformation matrix can be exactly decomposed in the form shown in Eq.~\eqref{eq:hermiticity2}. Otherwise, they coincide approximately. In the example provided in Sec.~\ref{sec:applications}, we use the construction of Eq.~\eqref{eq:chi_mode1}, since it is computationally easier to deal with to approximate the spectral density.

The coupling parameters are recovered from the eigenfrequencies $\tilde\omega_\mu$ by reversing the diagonalisation of the matrix $\mathbb H$, which appears in Eq~\eqref{eq:correlator_global_pseudomode}. Then we have that 
\begin{align}
    \omega_{\lambda\lambda'} &= (V^{-1}\tilde\omega V)^{\text{H}}_{\lambda\lambda'},\label{eq:coupling_h}\\
    \kappa_{\lambda\lambda'}-\gamma_{\lambda\lambda'}&= 2i(V^{-1}\tilde\omega V)^{\text{AH}}_{\lambda\lambda'}\label{eq:coupling_ah},
\end{align}
where $\tilde\omega$ is a diagonal matrix whose elements are the QNM's frequencies $\tilde\omega_\mu$, and the superscript $\text{AH}$ indicates the anti-Hermitian part. The parameters on the left-hand side of Eq.~\eqref{eq:coupling_ah} can be individually obtained by setting the positive eigenvalues of $2i(V^{-1}\tilde\omega V)^{\text{AH}}$ to be the eigenvalues of $\kappa$, and the modulus of the negative eigenvalues of $2i(V^{-1}\tilde\omega V)^{\text{AH}}$ to be the eigenvalues of the $\gamma$. This association ensures that $\kappa$ and $\gamma$ are positive semi-definite matrices, thus guaranteeing that the master equation~\eqref{eq:master} is physically meaningful \cite{Breuer2002_BOOK}.

\section{Example: One-dimensional slab cavity}
\label{sec:applications}
In this section, we apply and benchmark the method developed in this work for an explicit example geometry. More specifically, we use our approach to verify the quality of the approximation in Eq.~\eqref{eq:hermitized_pole_correlator}, that is, how well the matching condition is satisfied.

\subsection{Example geometry}

\subsubsection{Setup}

\begin{figure}[h]
    \centering
    \includegraphics[width = \columnwidth, trim={5cm 5cm 5cm 6.5cm}, clip]{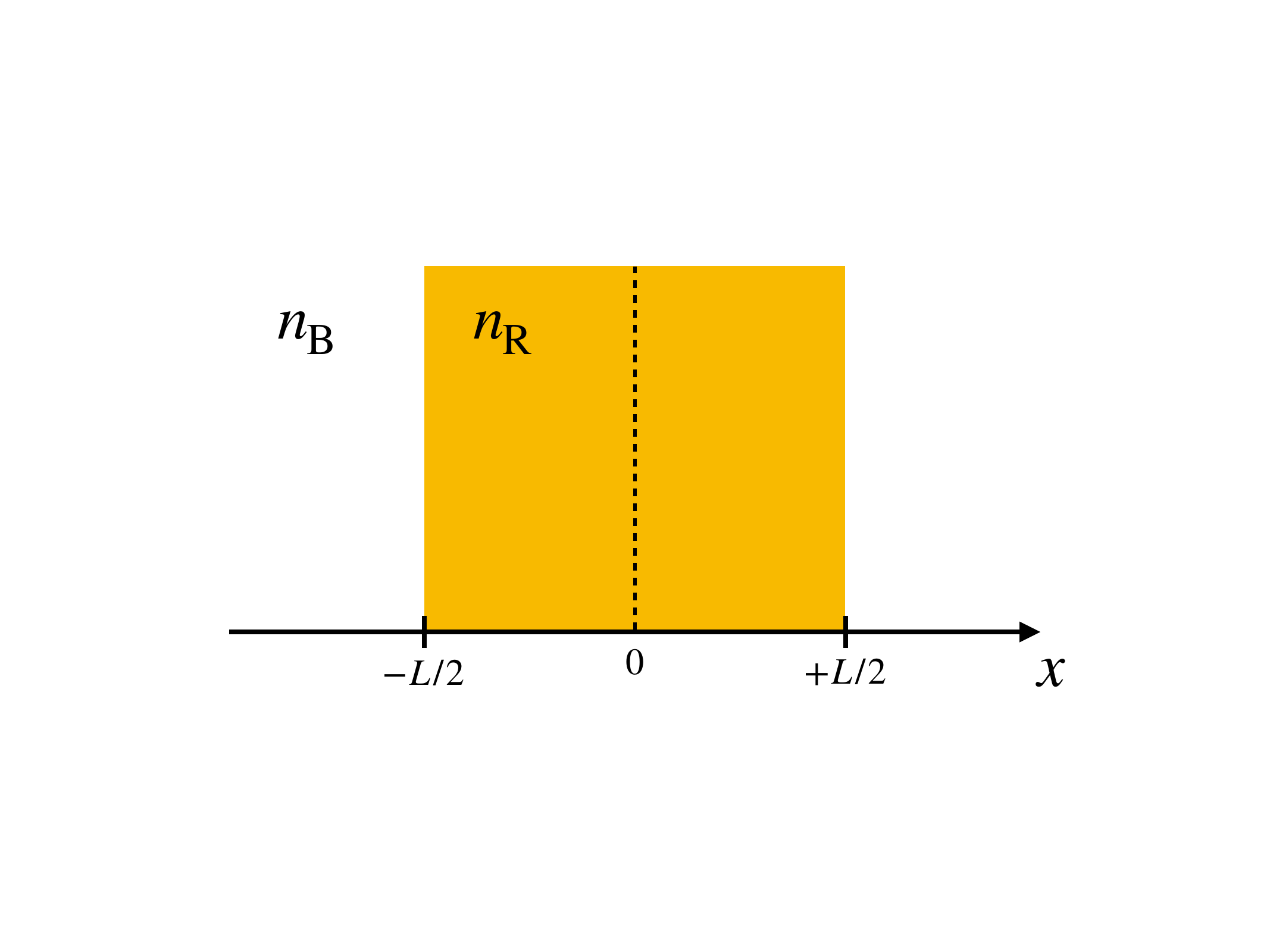}
    \caption{Setup of the one-dimensional example cavity~\cite{WeitzelThesis2025}. The horizontal axis indicates the position coordinate $x$, while $n_{\mathrm{R}}$ and $n_{\mathrm{B}}$ are the refractive indices of the slab and its surroundings, respectively.}
    \label{fig:fig_1-1D_example}
\end{figure}

We consider one of the simplest possible cases of a dielectric resonator: A one-dimensional, dispersionless and absorptionless cavity, as depicted in Fig.~\ref{fig:fig_1-1D_example}. It consists of a slab with constant refractive index $n_{\mathrm{R}}$ and length $L$, centered at the origin of the interval $[-L/2, L/2]$, and surrounded by a medium with refractive index $n_{\mathrm{B}}$. Both $n_{\mathrm{R}}$ and $n_{\mathrm{B}}$ are 
real, due to the absorptionless and dispersionless properties. We emphasize, however, that the formalism developed in this paper is not geometry-specific and can be applied to more general dispersionless dielectrics as well.

This setup is particularly useful as a testbed for the method, since many of its properties, including the quasi-normal modes, can be obtained analytically, as discussed in the following.

\subsubsection{Quasi-normal modes}

In the considered cavity geometry, the QNMs satisfy the one-dimensional Helmholtz equation \cite{Kristensen2020},
\begin{align}
    \left(\partial_x^2+n_{\mathrm{R}}^2\frac{\tilde\omega_\mu^2}{c^2}\right)\tilde f_\mu(x)=0,\label{eq:Helmholtz_equation_1d}
\end{align}
subject to the one-dimensional analog of the Silver-Müller radiation condition, Eq.~\eqref{eq:silver_mueller}, which reads
\begin{align}
    \lim_{x\to\infty}\tilde f_\mu(x)= \frac1{n_{\mathrm{B}}}\sqrt{\frac{\mu_0}{\epsilon_0}}\partial_x\tilde f_\mu(x).\label{eq:silver_mueller_1d}
\end{align}

The above equations can be solved analytically, and the QNM solution reads \cite{Kristensen2020}
\begin{align}
    \tilde f_\mu(x) &= A_\mu\left(e^{in_{\mathrm{R}}\tilde\omega_\mu x/c}+e^{-in_{\mathrm{R}}\tilde\omega_\mu x/c+i\mu\pi}\right) \,,\label{eq:QNMs_1D}
\end{align}
$\mu\in\mathbb Z$, with frequencies
\begin{align}
    \tilde\omega_\mu& =(\mu\pi+i\ln\alpha) \frac{c}{L n_{\mathrm{R}}} \,,\label{eq:freqs_QNMs_1D}
\end{align}
where $A_\mu = (e^{i\mu\pi/2}n_{\mathrm{R}}\sqrt{2L})^{-1}$ is a normalization factor and $\alpha\equiv |n_{\mathrm{R}}-n_{\mathrm{B}}|/(n_{\mathrm{R}}+n_{\mathrm{B}})$. We note that the Green's function for this geometry also admits an explicit analytical expression~\cite{Kristensen2020,Tomas1995,Lentrodt2022_pyrot_pypi}.

\subsection{Solution of the Hermitization condition}
We now apply our prescription for the construction of the gPM parameters, whose starting point is to solve the Hermitization condition, Eq.~\eqref{eq:hermiticity1}. Recall that only approximate solutions of the latter can be found, by truncating the solutions of Eq.~\eqref{eq:hermiticity_extended} to the block $\mu,\nu>0$. The latter, in turn, can be constructed directly from the QNMs, as shown in Eqs.~(\ref{eq:T1} --~\ref{eq:family}). 
Using the explicit formulas for the QNMs [Eq.~\eqref{eq:QNMs_1D}] and for their frequencies [Eq.~\eqref{eq:freqs_QNMs_1D}], we show, in Appendix~\ref{appendix::analytical_solution_T}, that an exact, although not Hermitian, solution of Eq.~\eqref{eq:hermiticity_extended} is
\begin{align}
    &\tilde{T}^{(\text{slab})}_{\mu\nu}=-\frac{2i\tilde\omega_\mu n_{\mathrm{R}}n_{\mathrm{B}}}{n_{\mathrm{R}}^2-n_{\mathrm{B}}^2}\frac{1+(-1)^{\mu-\nu}}{(\mu-\nu)\pi+2i\ln\alpha} \,.\label{eq:solution_1D_extended}
\end{align}
The above expression is found by choosing $a=2$ in the family of solutions [Eq.~\eqref{eq:family}]. We choose this value as it completely eliminates all anti-Hermitian terms of the form $\tilde\omega_{\mu}\delta_{\mu,-\nu}$.

Next, the approximate solution of Eq.~\eqref{eq:hermiticity1} is given by the Hermitian part of Eq.~\eqref{eq:solution_1D_extended}, which reads
\begin{align}
     \tilde T_{\mu\nu}^{(\text{slab}), \rm H}&= -\frac cL(\mu+\nu)\frac{\pi in_{\mathrm{B}}}{n_{\mathrm{R}}^2-n_{\mathrm{B}}^2}\frac{1+(-1)^{\mu-\nu}}{(\mu-\nu)\pi+2i\ln\alpha},\label{eq:Hermitian_approx}
\end{align}
but is restricted to the $\mu,\nu>0$ index domain. A more detailed comparison between Eq.~\eqref{eq:Hermitian_approx} and Eq.~\eqref{eq:solution_1D_extended} is carried out in Appendix~\ref{appendix::analytical_solution_T}. Essentially, the overall structure of Eq.~\eqref{eq:solution_1D_extended} is preserved by Eq.~\eqref{eq:Hermitian_approx}, and, thus, the latter is a suitable approximation of the former.
\begin{figure*}[t!]
    \centering
    \includegraphics[width=\textwidth]{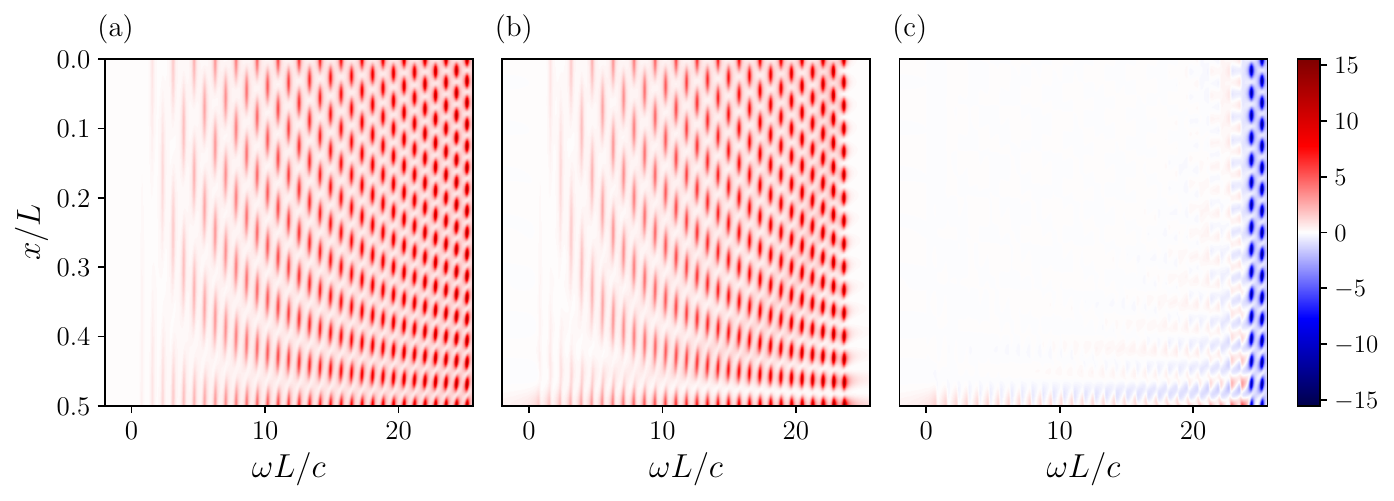}
    \caption{Spectral density for the one-dimensional slab cavity depicted in Fig.~\ref{fig:fig_1-1D_example}, for different frequencies and positions, considering $n_{\mathrm{R}}=4$ and $n_{\mathrm{B}}=1$ \cite{WeitzelThesis2025}. Only positions between the center ($x/L=0$) and the edge ($x/L=0.5$) of the cavity are shown, due to the symmetry of the latter. In (a), we show the exact spectral density obtained from the QNMs Green's function, via the correspondence in Eq.~\eqref{eq:correlator_to_green}. (b) depicts the spectral density obtained from a pole expansion constructed via Eq.~\eqref{eq:hermitized_pole_correlator_1d}, using $M=30$ terms in the summation. In (c), the difference between the spectral densities in (a) and (b) is shown. The colorbar is in units of $\hbar/(\epsilon_0 L)$.}
    \label{fig:fig_3-nR_4_2d_xw_comparison_exact_hermitized}
\end{figure*}

The next step is to verify whether the matrix in Eq.~\eqref{eq:Hermitian_approx} also satisfies Eq.~\eqref{eq:hermiticity2}, that is, if such matrix is positive semi-definite. We checked the latter property by numerically computing the associated eigenvalues, which were confirmed to be all real and positive.\footnote{Naturally, we computed the eigenvalues of the block associated with $\mu,\nu>0$ of Eq.~\eqref{eq:Hermitian_approx}.} Therefore, we can decompose the matrix Eq.~\eqref{eq:Hermitian_approx} as $\tilde T^{(\text{slab}), \rm H} = VV^\dagger$, and the Hermitization condition is satisfied. In our case, the matrix $V$ was constructed numerically from the eigenvectors and eigenvalues of $\tilde T^{(\text{slab}), \rm H}$. In the following, we use this solution for $V$ to test the resulting gPM representation of the spectral density for our example problem.

\subsection{Reproducing the position-resolved spectral density}
With our explicit expression in Eq.~\eqref{eq:Hermitian_approx}, we can construct an approximation for the spectral density. For the present setup, it reads as the one-dimensional version of Eq.~\eqref{eq:hermitized_pole_correlator} with $x=x'$, that is,
\begin{align}
    \label{eq:hermitized_pole_correlator_1d}
    \Im\tilde {C}^\mathrm{gPM}_{\mathrm{adv}}&(x, x, \omega)\nonumber\\=&\frac{\hbar}{2\epsilon_0}\Im\left[\sum_{\lambda,\lambda'}^M\chi_{\lambda}(x)(\mathbb H-\omega)^{-1}_{\lambda\lambda'}\chi^\dagger_{\lambda'}(x)\right] \,,
\end{align}
where $\chi_{\lambda}(x)$ and $\mathbb H$ are recovered from the QNMs via Eqs.~(\ref{eq:chi_mode1} --~\ref{eq:coupling_ah}). Furthermore, in Eq.~\eqref{eq:hermitized_pole_correlator_1d}, we explicitly chose an upper summation limit $M$, which correspond to the number of pseudomodes used in our calculations. Here and thereafter, we take $M=30$. 

We again recall that the importance of the expression in Eq.~\eqref{eq:hermitized_pole_correlator_1d} is to fulfill the constraints demanded by the gPMs ansatz, while complying with the matching condition, Eq.~\eqref{eq:matching_global_pseudomodes_to_green}. We can then check if the latter is satisfied by comparing Eq.~\eqref{eq:hermitized_pole_correlator_1d} with $\Im\tilde{C}^{\rm cont}_{\mathrm{adv}}(x, x,\omega)$, which, in turn, can be computed directly from the Green's function [see Eq.~\eqref{eq:correlator_to_green}].

We perform this comparison in Fig.~\ref{fig:fig_3-nR_4_2d_xw_comparison_exact_hermitized}. The figure shows the cavity field's spectral density as a function of frequency and position. Since the cavity shown in Fig.~\ref{fig:fig_1-1D_example} is symmetric, we restrict the plots to positions between $x/L=0$ and $x/L=0.5$, corresponding to the center and the edge of the cavity, respectively.

We see that there is an overall good agreement between the exact result and the gPM expansion. More specifically, the third panel shows that the difference between the two expansions is, generally, of a much smaller magnitude than the relevant values of both exact and approximate cases. The most noticeable deviations appear as two types of edge effects. The first is in the frequency domain, since we are only able to include a finite number $M$ of gPMs in the expansion. There is therefore a frequency cutoff, above which there are large deviations, in this case for frequencies higher than $\omega L/ c\approx 25$. Also for frequencies just below the cutoff region, there are growing deviations, since the excluded modes leak into this region. Physically, this effect is explained by the lossy nature of the considered cavity, which features overlapping modes.

The second edge effect is observed close to the spatial boundary of the cavity, at $x/L=0.5$. These deviations can be attributed to a slow convergence of Eq.~\eqref{eq:hermitized_pole_correlator_1d}, since the cavity edge also marks the boundary of the region of convergence of the Green's function pole expansion \cite{Leung1994,Kristensen2020}. Outside the cavity, the Green's function itself does not admit a pole expansion, and a treatment in this manner leads to a divergent expansion. We note, however, that leakage of photons to this region is physically included in the gPMs through the loss terms in the ansatz, as evidenced by the well-matching spectral density inside the resonator.

\begin{figure}
    \centering
    \includegraphics[width=\columnwidth]{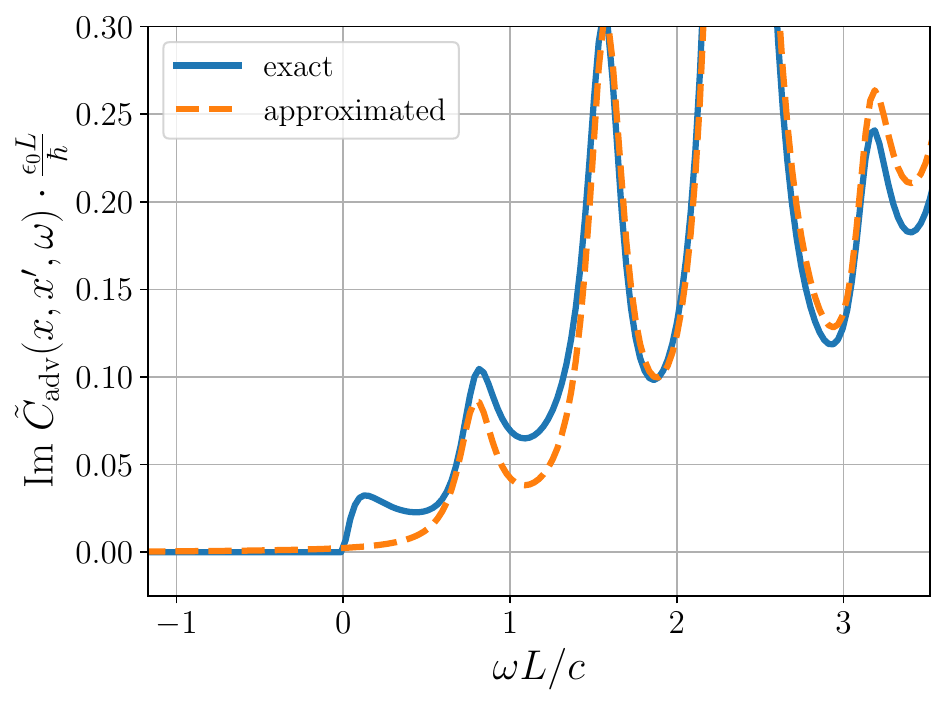}
    \caption{Spectral density for the one-dimensional example cavity depicted in Fig.~\ref{fig:fig_1-1D_example}, as a function of frequency, choosing $n_{\mathrm{R}}=4$ and $n_{\mathrm{B}}=1$ \cite{WeitzelThesis2025}. The plot corresponds to a horizontal cross section of the spectral densities in Fig.~\ref{fig:fig_3-nR_4_2d_xw_comparison_exact_hermitized} at position $x/L=0.15$, zooming into the region around $\omega L/c=0$. We used $M=30$ terms the expansion given by Eq.~\eqref{eq:hermitized_pole_correlator_1d}.}
    \label{fig:fig_4-spectral_density_plot}
\end{figure}

In order to see more clearly the implication of the meromorphic approximation, performed in Eq.~\eqref{eq:meromorphic_approx} by truncating the summation, Fig.~\ref{fig:fig_4-spectral_density_plot} shows the spectral density at a selected position, $x/L=0.15$. The figure thus corresponds to a horizontal cross section of Fig.~\ref{fig:fig_3-nR_4_2d_xw_comparison_exact_hermitized}, zoomed into the region around $\omega L/c=0$. We see the presence of deviations at low frequencies, which becomes more apparent at frequencies closer to $\omega L/c=0$. This feature arises because the truncation of the summation in Eq.~\eqref{eq:meromorphic_approx} discards the contribution of the poles in the left half of the complex plane, which particularly affects the low-frequency region. A related observation has also been made previously in standard pseudomodes theory \cite{Lednev2024}. Practically, it has been suggested that quantitative improvement can be achieved by complementing the gPMs with perturbative approaches to treat the residual contributions \cite{SanchezMartinez2024}. Non-physical extensions are also possible \cite{Lambert2019,Pleasance2020,Menczel2024}.

The above results therefore show that the gPMs method indeed provides a systematic method to fulfill the matching condition, even for highly lossy resonators with overlapping modes, as the one in Fig.~\ref{fig:fig_1-1D_example}. In Appendix~\ref{appendix:comparison_xx}, the performance of our method is further investigated as a function of both the position arguments of the correlator. That is, we fix a frequency $\omega$ and verify the quality of the approximation of Eq.~\eqref{eq:hermitized_pole_correlator} along the position coordinates $x$ and $x'$. Again, we conclude that our prescription leads to a very good approximation of the imaginary part of the correlator in the frequency domain. 

The good quality of the approximation of the spectral density using the gPM formalism, evidenced by the results presented in this section, implies that the dynamics of \textit{any} quantum system coupling to the cavity field is recovered. The continuum cavity field has therefore successfully been represented by discrete lossy modes. In Appendix~\ref{appendix:comparison_other_methods}, we further discuss how our approach performs in comparison with alternative methods that also provide the discretization of a continuous environment.

\section{Conclusions and outlook}

We introduced the method of global pseudomodes (gPMs) and applied it to the case of cavity quantum electrodynamics. The approach provides a systematic way to construct effective discrete cavity modes which mimic the effect of a continuous environment onto a quantum system. In particular, one of the central features in the gPMs formalism  was to explicitly incorporate the spatial coordinate in the computation of the free parameters of the model. This way, the model is not only able to simulate the dynamics of a quantum emitter at any position within the cavity, but also to account for the spatial extension of such emitters.

As a preliminary testbed, we investigated the performance of our method in the simple example of a one-dimensional and highly lossy cavity. In this setup, we found that the gPM construction is able to reproduce the field's exact spectral density with good agreement, both as a function of the frequency and of the position coordinate. This implies that our approach can indeed simulate the continuous electric field in a cavity. Consequently, it can reconstruct the expected time-evolution of a quantum system placed within the resonator. The investigation of this aspect will be left for future work.

Our formalism further opens multiple avenues to model quantum systems coupled to electromagnetic cavities. For example, as the gPMs incorporate the position degree of freedom in the formalism, they promise applications in the case of cQED with extended quantum systems, which are currently investigated particularly in solid-state \cite{Appugliese2022,Jarc2023} and many-body cavity QED \cite{Ritsch2013}. For these cases, the standard pseudomodes approach \cite{Garraway1997a, Garraway1997b, Dalton1999a} is insufficient, as it requires a numerical fitting in a (-n eventually) too large number of points along the position coordinate \cite{Lentrodt2020,Medina2021}. Our approach provides a different concept in this regard, since it removes the need of a fitting procedure.


We further envision extending our formalism to dispersive materials, as well as benchmarking our method with 3D resonators. The concept developed here in principle applies directly to the latter cases, apart from potentially needing modifications to incorporate dispersion in the formalism. However, solving the Hermitization condition may pose additional difficulties. Still, fitting procedures can again be useful, not directly for the correlator, as in \cite{Lambert2019,Medina2021}, but, instead, to numerically solve Eq.~\eqref{eq:matrix_squaree_root} for the matrix $V$.

\section{Acknowledgements}
The authors thank H.~P.~Breuer, J.~Brugger, E.~Brunner, J.~Evers, J.~Franz, M.~W.~Haverkort, F.~Riesterer and V.~Shatokhin for valuable discussions. DL gratefully acknowledges the Georg H. Endress Foundation for financial support. LW and DL gratefully acknowledge support by the DFG funded Research Training Group ``Dynamics of Controlled Atomic and Molecular Systems'' (RTG 2717).

\appendix

\section{Matching the two-point correlators of the field}
\label{appendix:matching condition}

In this Appendix, we formulate the correlator equivalence statement in its original form \cite{Menczel2024}, and show that it can be reduced to Eq.~\eqref{eq:matching_global_pseudomodes_to_green} in the main text, at least in the case where a quantum system couples with an electromagnetic environment in the cavity.

We start with the definitions of the advanced and retarded correlators of the electric field. They read, respectively \cite{Menczel2024}
\begin{align}
    \mathbf C_{\mathrm{adv}}(\vb r, \vb r', \tau)&\equiv i\langle\hat{\mathbf{E}}(\vb r, t+\tau) \otimes \hat{\mathbf{E}}(\vb r', t)\rangle,\label{eq:adv_correlator}\\
     \mathbf C_{\mathrm{ret}}(\vb r, \vb r', \tau)&\equiv i\langle\hat{\mathbf{E}}(\vb r, t) \otimes \hat{\mathbf{E}}(\vb r', t+\tau)\rangle\label{eq:ret_correlator},
\end{align}
where $t, \tau \geq 0$, and $\otimes$ denotes the dyadic product between two vectors. The $i$ factor is added for later convenience. The operator $\hat{\vb E}$ without any subscripts indicates the electric field either in the continuum or in the gPM representation. As in the main text, the expectation value is taken with respect to an decoupled initial state of the environment $\hat\rho_{\rm field}(0)$, that is $\langle\bullet\rangle = \Tr[\bullet\;\hat\rho_{\rm field}(0)]$, which is further required to be stationary. As also mentioned in the main text, the latter property makes the correlators~\eqref{eq:adv_correlator} and~\eqref{eq:ret_correlator} to be dependent only on the time interval $\tau$, so that we can set $t=0$ from now on.

The original equivalence statement from \cite{Menczel2024} then reads as follows: The time evolution of the reduced (matter) system state $\hat \rho_\mathrm{S}(t)$ obtained through the gPM ansatz [see Sec.~\ref{sec:gpm_setup}] is identical to the one obtained via the continuum representation for the fields if the correlators of the coupling operators in both representations match. Mathematically, if
\begin{align}
    \mathbf{C}^\mathrm{gPM}_{\mathrm{adv}}(\vb r,\vb r',\tau) \overset{!}{=} {\mathbf{C}}^\mathrm{cont}_{\mathrm{adv}}(\vb r, \vb r', \tau) \,, \label{eq::corr_match1}
    \\
    \mathbf{C}^\mathrm{gPM}_{\mathrm{ret}}(\vb r,\vb r',\tau) \overset{!}{=} {\mathbf{C}}^\mathrm{cont}_{\mathrm{ret}}(\vb r, \vb r', \tau) \,, \label{eq::corr_match2}
\end{align}
for all $\tau\geq 0$, and all positions $\vb r, \vb r'$ within the cavity volume. In addition, the stationary state of the field is required to be Gaussian, and the interaction Hamiltonian must have the bilinear structure already discussed in Sec.~\ref{sec:matching_condition} in the main text.

In order to show the equivalence between the statements of Eqs.~\eqref{eq::corr_match1} and~\eqref{eq::corr_match2}, and the matching condition of Eq.~\eqref{eq:matching_global_pseudomodes_to_green}, we have to suitably convert the advanced and retarded correlators to the frequency domain. To this end, we note that the correlators can also be thought of as parts of a general piecewise correlator, defined in the whole time domain as \cite{Menczel2024}
\begin{align}
    \vb C(\vb r, \vb r', \tau)=
    \begin{cases}
        i\langle\hat{\vb E}(\vb r, \tau)\otimes\hat{\vb E}(\vb r', 0)\rangle, &\tau\geq 0,\\
        i\langle\hat{\vb E}(\vb r, 0)\otimes\hat{\vb E}(\vb r', |\tau|)\rangle, &\tau< 0.
    \end{cases}
\end{align}
Since $\vb C(\vb r, \vb r', \tau)$ is defined for all times, we can compute its (full range) Fourier transform. It reads
\begin{align}
    \label{eq:correlator_fourier}
    \tilde{\vb C}(\vb r, \vb r', \omega) &= \int_0^\infty\dd \tau\;\vb C_{\mathrm{adv}}(\vb r, \vb r', \tau) e^{i\omega \tau}\nonumber\\&+\int_{0}^\infty\dd \tau\;\vb C_{\mathrm{ret}}(\vb r, \vb r', \tau) e^{-i\omega \tau} \,,
\end{align}
where, in the above, the Fourier transform was separated into two terms: The first is associated with the integration over positive times. The second term, associated to the integration over negative times, takes the form appearing in Eq.~\eqref{eq:correlator_fourier} after changing the integration variable as $\tau \to -\tau$.

The Fourier transform of the full correlator thus motivates us to define the one-sided Fourier transforms of advanced and retarded correlators as
\begin{align}
    \tilde{\vb C}_{\mathrm{adv}}(\vb r, \vb r', \omega)&\equiv \int_0^\infty\dd \tau\;\vb C_{\mathrm{adv}}(\vb r, \vb r', \tau) e^{i\omega \tau},\label{eq:advanced_correlator_spectrum_appendix}\\
    \tilde{\vb C}_{\mathrm{ret}}(\vb r, \vb r', \omega)&\equiv \int_0^\infty\dd \tau\;\vb C_{\mathrm{adv}}(\vb r, \vb r', \tau) e^{-i\omega \tau}\,,\label{eq:retarded_correlator_spectrum_appendix}
\end{align}
such that
\begin{align}
    \tilde{\vb C}(\vb r, \vb r', \omega) &=  \tilde{\vb  C}_{\mathrm{adv}}(\vb r, \vb r', \omega) + \tilde{\vb  C}_{\mathrm{ret}}(\vb r, \vb r', \omega) \,.
\end{align}

Therefore, we see that Eqs.~\eqref{eq::corr_match1} and~\eqref{eq::corr_match2} are equivalent, respectively, to
\begin{align}
    \tilde{\mathbf{C}}^\mathrm{gPM}_{\mathrm{adv}}(\vb r,\vb r',\omega) &\overset{!}{=} \tilde{\mathbf{C}}^\mathrm{cont}_{\mathrm{adv}}(\vb r, \vb r', \omega) \,, \label{eq::corr_match1_freq}
    \\
    \tilde{\mathbf{C}}^\mathrm{gPM}_{\mathrm{ret}}(\vb r,\vb r',\omega) &\overset{!}{=} \tilde{\mathbf{C}}^\mathrm{cont}_{\mathrm{ret}}(\vb r, \vb r', \omega) \, \label{eq::corr_match2_freq}.
\end{align}

Let us focus on the objects on the right-hand sides of Eqs.~\eqref{eq::corr_match1_freq} and~\eqref{eq::corr_match2}, that is, computed in the continuum representation of the field. We note that, in this representation, the advanced and retarded correlators are related by
\begin{align}
    \label{eq:unitary_env_corr}
    \vb C^{\rm cont}_\mathrm{adv}(\vb r, \vb r', t)=-[\vb C^{\rm cont}_\mathrm{ret}(\vb r, \vb r', t)]^*,
\end{align}
which is a general property satisfied by unitary environments \cite{Menczel2024}. The minus sign Eq.~\eqref{eq:unitary_env_corr} is due to the $i$ factor in the definitions in Eqs.~\eqref{eq:adv_correlator} and ~\eqref{eq:ret_correlator}. Hence, using the definitions of the one-sided Fourier transforms of Eqs.~\eqref{eq:advanced_correlator_spectrum_appendix} and~\eqref{eq:retarded_correlator_spectrum_appendix}, Eq.~\eqref{eq:unitary_env_corr} implies that \cite{WeitzelThesis2025}
\begin{align}
    \label{eq:rel1}
    \Im\tilde{\vb C}^{\rm cont}_{\mathrm{ret}}(\vb r, \vb r', \omega)&=\Im\tilde{\vb C}^{\rm cont}_{\mathrm{adv}}(\vb r, \vb r', \omega),\\
    \label{eq:rel2}
    \Re\tilde{\vb C}^{\rm cont}_{\mathrm{ret}}(\vb r, \vb r', \omega)&=-\Re\tilde{\vb C}^{\rm cont}_{\mathrm{adv}}(\vb r, \vb r', \omega).
\end{align}

In addition, it is known that every function constructed by one-sided Fourier transforms, as in Eqs.~\eqref{eq:advanced_correlator_spectrum_appendix} and~\eqref{eq:retarded_correlator_spectrum_appendix}, satisfies a Kramers-Kronig relation \cite{Nussenzveig1972, Scheel2008}. In the present context, it reads, for the advanced correlator
\begin{align}
    \label{eq:kramers-kronig}
    \Re\tilde {\vb C}^{\rm cont}_{\mathrm{adv}}(\vb r, \vb r', \omega) = \mathcal P\int_0^\infty\dd\omega'\frac{\Im\tilde {\vb C}^{\rm cont}_{\mathrm{adv}}(\vb r, \vb r', \omega')}{\omega'-\omega},\
\end{align}
with a similar relation holding for the retarded correlator, but with an overall minus sign on the left-hand side.

Due to Eqs.~\eqref{eq:rel1},~\eqref{eq:rel2} and~\eqref{eq:kramers-kronig}, we conclude that the continuum representation allows one to recover both the advanced and the retarded correlators solely from $\Im\tilde {\vb C}^{\rm cont}_{\mathrm{adv}}(\vb r, \vb r', \omega)$, by reverting the steps outlined above.

Turning now to the gPM representation, we show that the latter yields the very same constrains as Eqs.~\eqref{eq:rel1},~\eqref{eq:rel2} and~\eqref{eq:kramers-kronig}. To that end, we have to use the fact that the gPM ansatz allows for a pole expansion representation of $\Im\tilde {\vb C}^{\rm gPM}_{\mathrm{adv}}(\vb r, \vb r', \omega)$, as shown in Eq.~\eqref{eq:diag_propagator} in the main text. The proof of this result is provided in Appendix~\ref{appendix:correlator_pseudomodes}. Each term of this expansion therefore has the form of a complex Lorentzian function: $L(z) = r/(z-z_0)$, with $r,z_0\in\mathbb C$. Since it is known that the Lorentzian function fulfills the Kramers-Kronig relation \cite{Nussenzveig1972}, we conclude that
\begin{align}
    \label{eq:kramers-kronig_gpm}
    \Re\tilde {\vb C}^{\rm gPM}_{\mathrm{adv}}(\vb r, \vb r', \omega) = \mathcal P\int_0^\infty\dd\omega'\frac{\Im\tilde {\vb C}^{\rm gPM}_{\mathrm{adv}}(\vb r, \vb r', \omega')}{\omega'-\omega}.
\end{align}

In addition, since we require the gPM coupling operator --- i.e. the discrete mode expansion of the electric field in Eq.~\eqref{eq:electric_field_FM} in the main text --- to be a Hermitian operator, the relation in Eq.~\eqref{eq:unitary_env_corr} also holds for the gPMs, even though they constitute a non-unitary environment. Consequently, we have\footnote{In the light of Eq.~\eqref{eq:diag_propagator}, we further need to assume that $\overbar{\vb g}^*(\vb r)\otimes\tilde{\vb g}^{\rm T}(\vb r')=\tilde{\vb g}^*(\vb r)\otimes\overbar{\vb g}^{\rm T}(\vb r')$ \cite{WeitzelThesis2025}. Physically, the latter requirement is known as reciprocity, and it is already satisfied by the correlator in the continuum framework, due to its connection to the Green's function \cite{Buhmann2012}.}
\begin{align}
    \Im\tilde{\vb C}^{\rm gPM}_{\mathrm{ret}}(\vb r, \vb r', \omega)&=\Im\tilde{\vb C}^{\rm gPM}_{\mathrm{adv}}(\vb r, \vb r', \omega),\\
    \Re\tilde{\vb C}^{\rm gPM}_{\mathrm{ret}}(\vb r, \vb r', \omega)&=-\Re\tilde{\vb C}^{\rm gPM}_{\mathrm{adv}}(\vb r, \vb r', \omega).
\end{align}

Finally, we conclude that the advanced and retarded two-point correlators in the gPM ansatz are completely recovered by fixing  $\Im\tilde{\vb C}^{\rm gPM}_{\mathrm{adv}}(\vb r, \vb r', \omega)$, just like with the continuum representation. Therefore, since the correlators fulfill the exact same constrains in both frameworks, the conditions Eqs.~\eqref{eq::corr_match1} and~\eqref{eq::corr_match2} are completely equivalent to
\begin{align}
    \Im\tilde{\vb C}^{\mathrm{gPM}}_{\mathrm{adv}}(\vb r, \vb r', \omega) \overset{!}{=} \Im\tilde{\vb C}^{\mathrm{cont}}_{\mathrm{adv}}(\vb r, \vb r', \omega),
\end{align}
in order to guarantee that Eqs.~\eqref{eq::corr_match1} and~\eqref{eq::corr_match2} are satisfied.

\section{One-sided Fourier transform of correlators in the gPM framework}
\label{appendix:correlator_pseudomodes}

For the calculation of the advanced correlator using the gPM ansatz, we first work out the time evolution of the ladder operators $\hat a_\mu(t)$, \textit{decoupled} from the matter system. From Eq.~\eqref{eq:master}, one can show that the ladder operators in the Heisenberg picture evolve according to \cite{Breuer2002_BOOK},
\begin{align}
    &\partial_t{\hat a}_\mu(t)=\frac i\hbar [\hat H_\mathrm{gPM},\hat a_\mu(t)]\nonumber\\
    +&\sum_{\lambda,\lambda'}\kappa_{\lambda\lambda'}\left[\hat a_\lambda^\dagger(t) \hat a^{\strut}_\mu(t)\hat a^{\strut}_{\lambda'}(t)-\frac12\{\hat a_\lambda^\dagger(t) \hat a^{\strut}_{\lambda'}(t),\hat a^{\strut}_\mu(t)\}\right]\nonumber\\
    +&\sum_{\lambda,\lambda'}\gamma_{\lambda\lambda'}\left[\hat a^{\strut}_\lambda(t) \hat a^{\strut}_\mu(t)\hat a^\dagger_{\lambda'}(t)-\frac12\{\hat a^\dagger_\lambda(t) \hat a^{\strut}_{\lambda'}(t), \hat a^{\strut}_\mu(t)\}\right],\label{eq:adjoint_master_appendix}
\end{align}
where the gPM free Hamiltonian in the commutator is given by Eq.~\eqref{eq:hamiltonian_pseudomodes}. Employing the canonical commutation relations between the ladder operators given by  Eqs.~\eqref{eq:canonical_commutation} and~\eqref{eq:canonical_commutation2}, the master equation straightforwardly reduces to
\begin{align}
    \partial_t\hat a_\lambda(t)=-i\sum_{\lambda'}\mathbb H_{\lambda\lambda'}\hat a_{\lambda'}(t),\label{eq:master_simplified}
\end{align}
where $\mathbb H$ is the matrix of coefficients $\mathbb H_{\lambda\lambda'}=\omega_{\lambda\lambda'}-i(\kappa_{\lambda\lambda'}-\gamma_{\lambda\lambda'})/2$.

To find an explicit expression for the one-sided Fourier transform of the advanced  correlator [see Eq.~\eqref{eq:advanced_correlator_spectrum_appendix}] in the gPM representation, we first define
\begin{align}
    \mathcal C^{\mathrm{adv}}_{\lambda\lambda'}(t)=\mel{0}{\hat a^{\strut}_\lambda(t)\hat a^\dagger_{\lambda'}(0)}{0},\;\;t\geq 0.\label{eq:correlator_ladder}
\end{align}
Also, the half-range Fourier transform allows a natural conversion to the Laplace transform by redefining the frequency variable as $i\omega\to -s$. We denote the Laplace transform as
\begin{align}
    \overbar{\mathcal C}^{\mathrm{adv}}_{\lambda\lambda'}(s)\equiv \int_0^{+\infty}\dd t\;e^{-st}\mathcal C^{\mathrm{adv}}_{\lambda\lambda'}(t).
\end{align}

In order to proceed, we take the time derivative of $\mathcal C^{\mathrm{adv}}_{\lambda\lambda'}(t)$ and directly\footnote{Formally, this procedure is guaranteed by the quantum regression theorem \cite{Breuer2002_BOOK}.} employ the result of Eq.~\eqref{eq:master_simplified}. We thus get a differential equation for the correlator for positive times,
\begin{align}
    \partial_t\mathcal C^{\mathrm{adv}}_{\lambda\lambda'}(t)=-i\sum_{\eta}\mathbb H_{\lambda\eta}\mathcal C^{\mathrm{adv}}_{\eta\lambda'}(t),\label{eq:correlator_equation}
\end{align}
which can be converted to an algebraic equation by taking its Laplace transform, that is
\begin{align}
    s\overbar{\mathcal C}^{\mathrm{adv}}_{\lambda\lambda'}(s)-\mathcal C^{\mathrm{adv}}_{\lambda\lambda'}(0)=-i\sum_{\eta}\mathbb H_{\lambda\eta}\overbar{\mathcal C}^{\mathrm{adv}}_{\eta\lambda'}(s).
\end{align}
Next, one can rearrange the terms above to get
\begin{align}
    \overbar{\mathcal C}^{\mathrm{adv}}_{\lambda\lambda'}(s)=\sum_{\eta}(s+i\mathbb H)^{-1}_{\lambda\eta}\mathcal C^{\mathrm{adv}}_{\eta\lambda'}(0).
\end{align}

We obtain the correlator in the frequency domain upon substituting $s=-i\omega$. Provided that the initial condition is $\mathcal C^{\mathrm{adv}}_{\eta\lambda}(0)=\delta_{\eta\lambda}$, since we are assuming the pseudomodes to be initially in the vacuum state, we find
\begin{align}
    \tilde{\mathcal C}^{\mathrm{adv}}_{\lambda\lambda'}(\omega)=i(\omega-\mathbb H)^{-1}_{\lambda\lambda'}.\label{eq:resolvent_advanced}
\end{align}

The expression for the one-sided Fourier transform of the advanced correlator is then found by employing the gPM expansion in Eq.~\eqref{eq:electric_field_FM}. This yields
\begin{align}
    &\tilde{\vb C}^{\rm gPM}_{\mathrm{adv}}(\vb r,\vb r',\omega)\nonumber\\&=i\int_0^\infty\dd t \;e^{i\omega t}\sum_{\lambda,
    \lambda'}\boldsymbol\chi_{\lambda}(\vb r)\otimes\boldsymbol\chi^{\dagger}_{\lambda'}(\vb r')\mel{0}{\hat a_\lambda(t)\hat a^\dagger_{\lambda'}(0)}{0} \,,
\end{align}
where the integration over the time domain corresponds precisely to the result of Eq.~\eqref{eq:resolvent_advanced}. Finally, we find
\begin{align}
    \tilde{\vb C}^{\rm gPM}_{\mathrm{adv}}(\vb r,\vb r',\omega)=\frac{\hbar}{2\epsilon_0}\sum_{\lambda,
    \lambda'}\boldsymbol\chi_{\lambda}(\vb r)(\mathbb H-\omega\mathbb I)^{-1}_{\lambda\lambda'}\boldsymbol\chi^\dagger_{\lambda'}(\vb r') \,,\label{eq:correlator_global_pseudomode_appendix} 
\end{align}
which is Eq.~\eqref{eq:correlator_global_pseudomode} in the main text.

\section{Pole expansion of $\omega^2\vb G(\vb r,\vb r',\omega)$}
\label{appendix:Green_pole_expansion}

The Green's function admits a pole expansion over the QNMs $\tilde{\vb f}_\mu(\vb r)$ as \cite{Kristensen2020}
\begin{align}
    \vb G(\vb r, \vb r',\omega)=\frac{c^2}{2\omega}\sum_{\mu=-\infty}^{+\infty}\frac{\tilde{\vb f}_\mu(\vb r)\otimes\tilde{\vb f}_\mu(\vb r')}{\tilde\omega_\mu-\omega} \,.\label{eq:green_qnm}
\end{align}
From Eq.~\eqref{eq:green_qnm}, we can straightforwardly calculate the residues of $\omega^2\vb G(\vb r, \vb r',\omega)$ at each pole $\tilde\omega_\mu$:
\begin{align}
    \text{Res}[\omega^2\vb G(\vb r, \vb r',\omega), \tilde\omega_\mu]=-\frac{c^2}{2}\tilde\omega_\mu\tilde{\vb f}_\mu(\vb r)\otimes\tilde{\vb f}_\mu(\vb r').
\end{align}
The above expression for the residues can be used in the Mittag-Leffler expansion theorem, which provides a way to construct a pole expansion of a meromorphic function~\cite{Ablowitz2003}. We find that
\begin{align}
    \omega^2\vb G(\vb r,\vb r',\omega)&=\frac{c^2}{2}\sum_{\mu=-\infty}^{+\infty}\left(\frac1{\tilde\omega_\mu-\omega}-\frac1{\tilde\omega_\mu}\right)\tilde\omega_\mu\tilde{\vb f}_\mu(\vb r)\otimes\tilde{\vb f}_\mu(\vb r')\nonumber\\
    &=\boldsymbol{\Delta}(\vb r,\vb r')+\frac{c^2}{2}\sum_{\mu=-\infty}^{+\infty}\frac{\tilde\omega_\mu\tilde{\vb f}_\mu(\vb r)\otimes\tilde{\vb f}_\mu(\vb r')}{\tilde\omega_\mu-\omega} \,.\label{eq:pole_green_shift}
\end{align}
In Eq.~\eqref{eq:pole_green_shift}, there are two terms. The first one is defined as
\begin{align}
    \boldsymbol{\Delta}(\vb r,\vb r') \equiv \frac{c^2}2\sum_{\mu=-\infty}^{+\infty} \tilde{\vb f}_\mu(\vb r)\otimes\tilde{\vb f}_\mu(\vb r') \,,
\end{align}
while the second corresponds to the actual expansion over the poles. Although the sums in each of the individual terms diverge, their combination does not. We note, though, that $\boldsymbol\Delta(\vb r,\vb r')$ is not only frequency-independent but also a purely real function, which can be deduced from the properties of the QNMs discussed in Appendix~\ref{appendix:qnm_properties}. Consequently, only the real part of the second term on the right-hand side of Eq.~\eqref{eq:pole_green_shift} diverges. Nevertheless, this divergence does not affect our method, since the matching condition Eq.~\eqref{eq:matching_global_pseudomodes_to_green} involves only the \textit{imaginary} part of $\omega^2\vb G(\vb r,\vb r',\omega)$, via Eq.~\eqref{eq:correlator_to_green}, and the expansion in Eq.~\eqref{eq:green_qnm2} is finite.

For the particular case of the dielectric slab considered in Sec.~\ref{sec:applications}, we remark that we can use the completeness property of the QNMs [Eq.~\eqref{eq:completeness}] to show that
\begin{align}
    \Delta(x,x')=\frac{c^2}{n_{\mathrm{R}}^2}\delta(x-x').
\end{align}
In other words, the real part of Eq.~\eqref{eq:pole_green_shift} diverges in a delta function-like form, i.e., only at the coincidence $x=x'$. This divergence is well-known and can be renormalized \cite{Buhmann2012}. For the case of cavities featuring absorption, also the imaginary part diverges such that so-called local-field corrections may be required \cite{Buhmann2012}.

\section{Useful properties of the quasi-normal modes and solving the Hermitization condition}
\label{appendix:qnm_properties}

In this Appendix, we outline some of the most relevant properties of the QNMs. Thereafter, we use these properties to solve the Hermitization condition [Eq.~\eqref{eq:matrix_squaree_root}].

The QNMs satisfy the following relations:
\begin{itemize}
    \item[(i)] The complex conjugate of a QNM also belongs to the set of QNMs, fulfilling the relation $\tilde{\vb f}_\mu^*(\vb r)=\tilde{\vb f}_{-\mu}(\vb r)$. The corresponding frequencies also satisfy a conjugation relation, given by $\tilde\omega^*_{\mu}=-\tilde\omega_{-\mu}$.

    \item[(ii)] For dielectric dispersionless materials, the QNMs satisfy a completeness relation,\footnote{It is not known, in the general case, for which domain of $\vb r$ and $\vb r'$ the completeness relation holds. It is conjectured to be restricted to the resonator volume, which has been proven to be the case for the special cases of 1D cavities \cite{Leung1994} and a 3D spherical cavity \cite{Kristensen2020}.} given by
    \begin{align}
        \frac12&\sum_{\mu=-\infty}^{+\infty} n^2(\vb r)\tilde{\vb f}_\mu(\vb r)\otimes\tilde{\vb f}_\mu(\vb r')=\vb I\delta(\vb r-\vb r') \,.\label{eq:completeness}
    \end{align}

    \item[(iii)] The QNMs are \textit{linearly dependent} to each other, as expressed by
    \begin{align}
        \sum_{\mu=-\infty}^{+\infty}\frac1{\tilde\omega_\mu}\tilde{\vb f}_\mu(\vb r)\otimes\tilde{\vb f}_\mu(\vb r')=0,\label{eq:overcompleteness}
    \end{align}
    meaning that the QNMs actually form an \textit{overcomplete} set.
\end{itemize}

We can use properties (i)--(iii) above to arrive at a solution for Eq.~\eqref{eq:hermiticity_extended}, which we repeat below for clarity
\begin{align}
    \tilde\omega_\mu\tilde{\vb f}_{\mu}(\vb r)=\sum_{\nu = -\infty}^{\infty} \tilde{T}_{\mu\nu}\tilde{\vb f}^*_{\nu}(\vb r),\label{eq:hermiticity_extended_appendix}
\end{align}
and encompasses the extended index domain. As already mentioned in the main text, we need to extend the index domain because properties (ii) and (iii) involve \textit{all} QNMs. The solution of Eq.~\eqref{eq:hermiticity_extended_appendix} will then serve as an approximate solution of the actual Hermitization condition, which encompasses the restricted index domain of $\mu,\nu>0$.

First, we note that Eq.~\eqref{eq:hermiticity_extended_appendix} can also be interpreted as writing $\tilde\omega_\mu\tilde{\vb f}_{\mu}(\vb r)$ as a linear combination of $\tilde{\vb f}^*_{\mu}(\vb r)$. Due to the overcompleteness of the QNMs, pointed out by property (iii), we see that multiple solutions for the matrix $\tilde T$ are allowed. One of such solutions is trivially obtained from property (i), and reads
\begin{align}
    \label{eq:T1_appendix}
    \tilde T_{\mu\nu}^{(1)} = \tilde\omega_\mu\delta_{\mu,-\nu},
\end{align}
which is Eq.~\eqref{eq:T1} in the main text. We note that, using the conjugation property of the QNM frequencies pointed by property (i), $\tilde T_{\mu\nu}^{(1)}=-\tilde T_{\nu\mu}^{(1)*}$. In other words, this solution is completely anti-Hermitian.

A second, independent, solution is found by employing property (ii) as follows,
\begin{align}
    &\tilde\omega_\mu\tilde{\vb f}_{\mu}(\vb r) = \tilde\omega_\mu\int_{\mathcal R}\dd^3\vb r'\;\tilde{\vb f}_{\mu}(\vb r')\delta(\vb r'-\vb r)\\
    &=\frac{\tilde\omega_\mu}2\sum_{\nu=-\infty}^{+\infty}\left[\int_{\mathcal R}\dd^3\vb r' n^2(\vb r')\tilde{\vb f}_{\mu}(\vb r')\cdot\tilde{\vb f}^*_\nu(\vb r')\right]\tilde{\vb f}^*_\nu(\vb r),\label{eq:proof_t2}
\end{align}
where $\mathcal R$ denotes the cavity volume, and we used the property that $n(\vb r)\in\mathbb R$, as the cavity material is assumed to be absorptionless and dispertionless. Therefore, from Eq.~\eqref{eq:proof_t2}, we arrive at
\begin{align}
    \tilde{T}^{(2)}_{\mu\nu}=\frac{\tilde\omega_\mu}2\int_{\mathcal R}\dd^3\vb r\;n^2(\vb r)\tilde{\vb f}_\mu(\vb r)\cdot\tilde{\vb f}^*_\nu(\vb r),\label{eq:T2_appendix}
\end{align}
which corresponds to Eq.~\eqref{eq:T2} in the main text.

Finally, we arrive at the family of solutions of Eq.~\eqref{eq:family} by a suitable linear combination of the solutions in Eq.~\eqref{eq:T1_appendix} and~\eqref{eq:T2_appendix}, as
\begin{align}
    \tilde T_{\mu\nu} = (1-a)\tilde{T}^{(1)}_{\mu\nu}+a\tilde{T}^{(2)}_{\mu\nu},\;a\in\mathbb C.
\end{align}

The solutions encompassed by the above family all satisfy Eq.~\eqref{eq:hermiticity_extended_appendix}, and are used in this work as approximate solutions of Eq.~\eqref{eq:hermiticity1}. However, as we mention in the text, we choose an $a$ which sets all anti-Hermitian terms of the form of $\tilde{T}^{(1)}_{\mu\nu}$ to zero. Such a choice is an important step to find an approximate solution that complies with the constrain imposed Eq.~\eqref{eq:hermiticity2}. Although choosing $a=1$ indeed sets the explicit term with $\tilde{T}^{(1)}_{\mu\nu}$ to zero, it is not guaranteed to do the same with the \textit{implicit} contributions from $\tilde{T}^{(2)}_{\mu\nu}$. We leave the derivation of a suitable $a$ to Appendix~\ref{appendix::analytical_solution_T}, where we turn to the specific cavity example discussed in Sec.~\ref{sec:applications}.

\section{Approximate analytical solution of the Hermitization condition for the 1D example cavity}\label{appendix::analytical_solution_T}
\begin{figure*}
    \centering
    \includegraphics[width=0.8\textwidth]{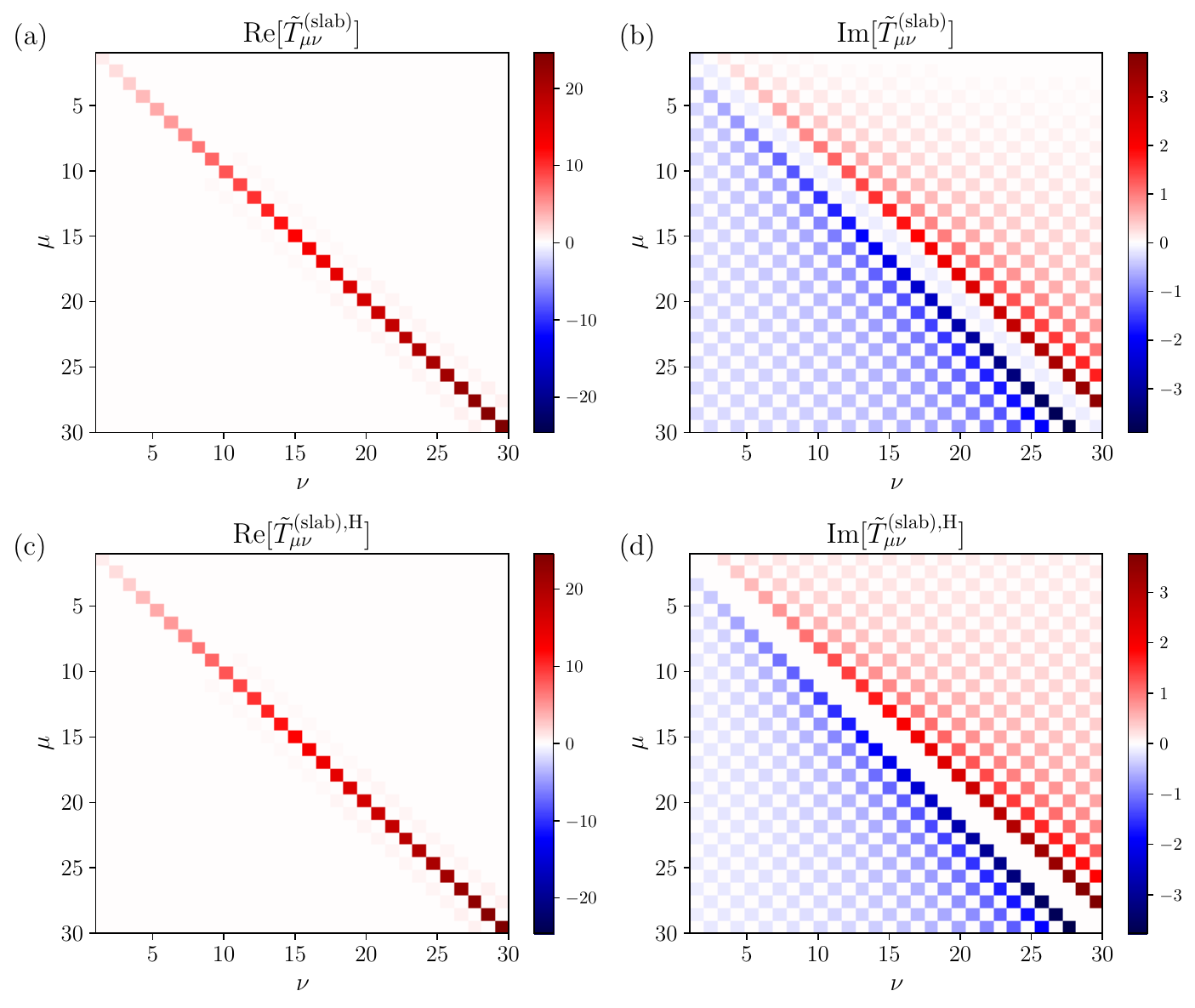}
    \caption{Block corresponding to $\mu,\nu>0$ of the matrix solution $\tilde T^{\rm (slab)}_{\mu\nu}$, given by Eq.~\eqref{eq:solution_1D}, and of its Hermitian part $\tilde T^{\rm (slab),\,H}_{\mu\nu}$, given by Eq.~\eqref{eq:solution_1d_hermitian}~\cite{WeitzelThesis2025}. (a) and (b), respectively, show the real and imaginary parts of the former, while (c) and (d) show the real and imaginary parts of the latter. The parameters $n_{\mathrm{R}}=4$ and $n_{\mathrm{B}}=1$ were chosen. The colorbar indicates the numerical value of each matrix element, in units of $c/L$.}
    \label{fig:fig_2-Hermitian_transformation}
\end{figure*}
\begin{figure*}[t]
    \centering
    \includegraphics[width=\textwidth]{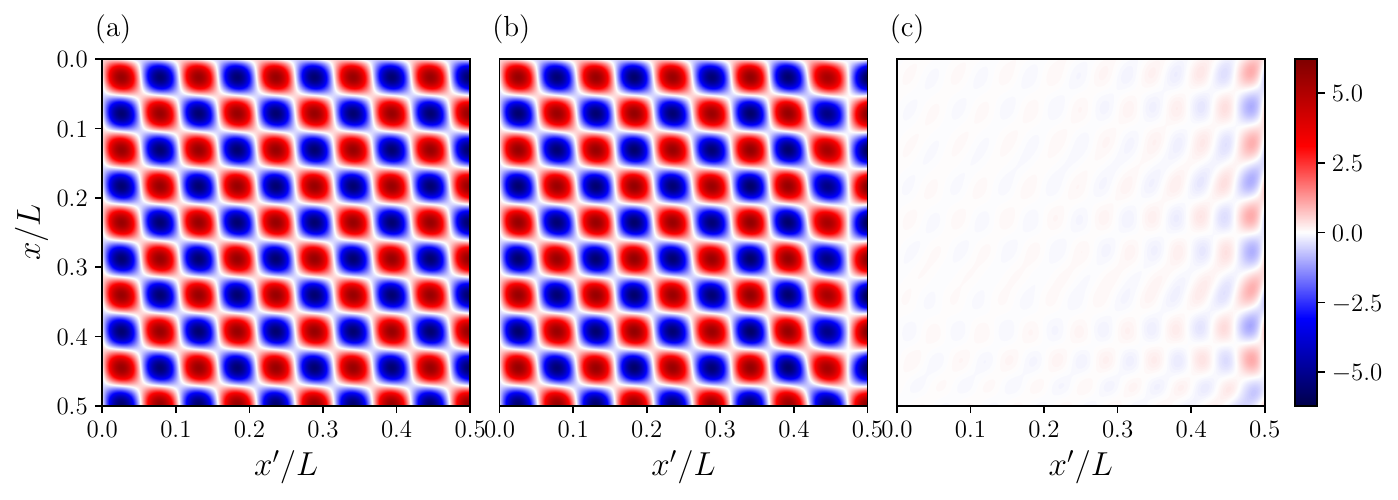}
    \caption{Imaginary part of the one-sided Fourier transform of the advanced correlator for the one-dimensional cavity slab depicted in Fig.~\ref{fig:fig_1-1D_example}, for different position arguments $x$, $x'$, and a fixed frequency \cite{WeitzelThesis2025}. We consider $n_{\mathrm{R}}=4$, $n_{\mathrm{B}}=1$ and set $\omega L/c = 15$.
    In (a), we depict the exact computation from the Green's function, via Eq.~\eqref{eq:correlator_to_green}; in (b), the approximation via Eq.~\eqref{eq:hermitized_pole_correlator_1d}, using $M=30$ terms, and, in (c), the difference between the two previous plots. The colorbar is in units of $\hbar/(\epsilon_0 L)$.}
    \label{fig:fig_5-nR_4_2d_xy_comparison_exact_hermitized}
\end{figure*}

In the following, we provide details on the calculations performed to arrive at Eq.~\eqref{eq:Hermitian_approx}. As discussed in the main text, we consider exact solutions of Eq.~\eqref{eq:hermiticity_extended} in the extended index domain. First, the trivial solution reads
\begin{align}
    \tilde T^{(\mathrm{slab}, 1)}_{\mu\nu}=\tilde\omega_\mu\delta_{\mu,-\nu},
\end{align}
where the frequencies $\tilde\omega_\mu$ are given by Eq.~\eqref{eq:freqs_QNMs_1D}. Next, we calculate the other possible transformation matrix by plugging into Eq.~\eqref{eq:T2} the QNMs given by Eq.~\eqref{eq:QNMs_1D} in the main text, that is, we evaluate
\begin{align}
    \tilde T^{(\text{slab},2)}_{\mu\nu} = \frac{\tilde\omega_\mu}2\int_{-L/2}^{+L/2}\dd x\;n^2_R\tilde{f}_\mu(x)\tilde{f}^*_\nu(x).\label{eq:trivial_1d}
\end{align}
Performing the integration over the position interval $[-L/2, L/2]$ leads to
\begin{align}
    \tilde T^{(\text{slab},2)}_{\mu\nu}=\frac{\tilde\omega_\mu}{2}\left[\delta_{\mu,-\nu}-\frac{2in_{\mathrm{R}}n_{\mathrm{B}}}{n_{\mathrm{R}}^2-n_{\mathrm{B}}^2}\frac{1+(-1)^{\mu-\nu}}{(\mu-\nu)\pi+2i\ln\alpha}\right],\label{eq:transformation_1d}
\end{align}
where $\alpha\equiv |n_{\mathrm{R}}-n_{\mathrm{B}}|/(n_{\mathrm{R}}+n_{\mathrm{B}})$.

The family of solutions is subsequently constructed by a linear combination [see Eq.~\eqref{eq:family}] of the solutions in Eqs.~\eqref{eq:trivial_1d} and \eqref{eq:transformation_1d}. We find
\begin{align}
    &\tilde T^{(\text{slab})}_{\mu\nu}=(1-a)\tilde\omega_\mu\delta_{\mu,-\nu}\nonumber\\&+\frac{a\tilde\omega_\mu}{2}\left[\delta_{\mu,-\nu}-\frac{2in_{\mathrm{R}}n_{\mathrm{B}}}{n_{\mathrm{R}}^2-n_{\mathrm{B}}^2}\frac{1+(-1)^{\mu-\nu}}{(\mu-\nu)\pi+2i\ln\alpha}\right].\label{eq:family_1d}
\end{align}
where $a$ is a free parameter. In order for the matrix to satisfy the constraints of the Hermitization condition --- that is, to be decomposable as $T = VV^\dagger$ --- it must be Hermitian. Since the term $\tilde\omega_\mu\delta_{\mu,-\nu}$ is completely anti-Hermitian, we will choose $a$ such that this term is eliminated. In other words, we select the ``most Hermitian" solution within the family in Eq.~\eqref{eq:family_1d}. Notice that the choice of $a=2$ allows the term $\tilde\omega_\mu\delta_{\mu,-\nu}$ to be completely eliminated, and we arrive at
\begin{align}
    &\tilde{T}^{(\text{slab})}_{\mu\nu}\big|_{a=2}=-\frac{2i\tilde\omega_\mu n_{\mathrm{R}}n_{\mathrm{B}}}{n_{\mathrm{R}}^2-n_{\mathrm{B}}^2}\frac{1+(-1)^{\mu-\nu}}{(\mu-\nu)\pi+2i\ln\alpha} \,.\label{eq:solution_1D}
\end{align}
While the solution in Eq.~\eqref{eq:solution_1D} is still exact in the extended index domain, it is still not Hermitian. Thus, to comply with the Hermitization condition, we approximate $\tilde{T}^{(\text{slab})}_{\mu\nu}\big|_{a=2}$ by its own Hermitian part
\begin{align}
    \tilde T^{(\text{slab})}_{\mu\nu}\approx &\frac12\left(\tilde{T}^{(\text{slab})}_{\mu\nu}+\tilde{T}^{(\text{slab}), \mathrm{H}}_{\mu\nu}\right).\label{eq:approximation_1d}
\end{align}
Finally, the approximate solution to Eq.~\eqref{eq:hermiticity1} is found by restricting the matrix in Eq.~\eqref{eq:approximation_1d} to the block where $\mu,\nu>0$. The resulting matrix is the one shown in Eq.~\eqref{eq:Hermitian_approx} of the main text, that is
\begin{align}
    \label{eq:solution_1d_hermitian}
     \tilde T_{\mu\nu}^{(\text{slab}), \mathrm H}&= -\frac cL(\mu+\nu)\frac{\pi in_{\mathrm{B}}}{n_{\mathrm{R}}^2-n_{\mathrm{B}}^2}\frac{1+(-1)^{\mu-\nu}}{(\mu-\nu)\pi+2i\ln\alpha} \,.
\end{align}

In Fig.~\ref{fig:fig_2-Hermitian_transformation} we show how the approximation in Eq.~\eqref{eq:Hermitian_approx} affects the block corresponding to $\mu,\nu>0$ of the exact matrix of Eq.~\eqref{eq:solution_1D_extended}. The figure shows that the latter equation conserves the general structure of the former. Namely, in both cases, the ``checkerboard" pattern of the matrix --- caused by the $1+(-1)^{\mu-\nu}$ factor --- is preserved, and in their real parts the non-zero elements are concentrated in and around the main diagonal. However, one sees that the imaginary part depicts more visible differences: The main diagonal of Eq.~\eqref{eq:solution_1D} has a non-zero imaginary contribution, which is eliminated by taking its Hermitian part. Also, we observe that the imaginary part in the non-Hermitian case is asymmetric with respect to the main diagonal, compared with the Hermitian case, which is symmetric by construction. Nevertheless, these differences occur on a much smaller scale compared to the absolute values of the matrix elements, as it can be seen from the color code in Fig.~\eqref{fig:fig_2-Hermitian_transformation}. Therefore, we conclude that Eq.~\eqref{eq:Hermitian_approx} well approximates Eq.~\eqref{eq:solution_1D_extended} and, hence, provides a good approximate solution to Eq.~\eqref{eq:hermiticity1}.

\section{Comparison of the correlator for a fixed frequency and different position arguments}
\label{appendix:comparison_xx}

The main strength of the gPMs lies in the property of well-approximating the spectral density throughout the cavity region, which promises applications for extended systems coupling to resonators. In Fig.~\ref{fig:fig_5-nR_4_2d_xy_comparison_exact_hermitized}, we extend the comparison of the imaginary part of the half-range Fourier transform of the advanced correlator to all accessible position arguments $x, x'$, while choosing a fixed frequency, $\omega L/c = 4$. We see that the agreement is good for almost all position arguments: The order of magnitude of the differences between the two computed objects is generally much smaller than the relevant values those objects take. However, there is a more significant disagreement alongside the position coordinate $x$, when the other coordinate is close to the edge of the cavity, that is $x'/L\approx 0.5$. This mismatch stems from the slow convergence of the pole expansion near the cavity's boundary. Interestingly, the disagreement is not symmetrical upon swapping $x$ with $x'$. This feature, in turn, is attributed to the fact that the matrix $\tilde T^{\rm (slab), H}_{\mu\nu}$, used in the construction of Eq.~\eqref{eq:hermitized_pole_correlator_1d}, is only an approximate solution to the Hermitization condition [Eq.~\eqref{eq:matrix_squaree_root}]. Hence, the construction of the gPMs via Eqs.~\eqref{eq:chi_mode1} and~\eqref{eq:chi_mode2} are not equivalent anymore, and choosing one construction --- in our case, Eqs.~\eqref{eq:chi_mode1} --- leads to the visible asymmetric mismatch, along only one of the position coordinates.

\section{Comparison to alternative methods}
\label{appendix:comparison_other_methods}

From a general perspective, the gPMs are a type of few-mode theory for cavity QED --- that is an approach which describes the cavity field by discrete modes instead of a continuum \cite{Lentrodt2020,Medina2021}. There have been various such approaches in the literature with different advantages, with a few of them surveyed in the introduction (Sec.~\ref{sec:intro}). Our approach is mostly based on pseudomode theory~\cite{Garraway1997a,Garraway1997b,Tamascelli2018,Lambert2019,Menczel2024} and extends it to a position-resolved description. The concept of gPMs is set up to fulfill strict physical conditions and a one-to-one correspondence between modes and resonances. This makes it rather appealing to subsequently solve problems where quantum systems couple to the cavity. There are, however, other approaches which feature similar constructions. In this Appendix, we compare a selection of these methods to the gPMs by checking how well they approximate the cavity field's spectral density

An approach which has received much attention recently are the so-called \textit{quantized quasi-normal modes}. Originally introduced in \cite{Franke2019} and further developed in \cite{Franke2020a, Franke2020b, Franke2023}, this method attempts to establish an \textit{ab initio} quantization of the resonant states, as the name suggests. To this end, the quasi-normal modes are associated with quantum operators, and a subsequent manipulation of their dynamical equations leads to a master equation in Lindblad form. Both the final master equation and an associated mode transformation --- which the authors call symmetrization transformation \cite{Franke2019} --- are rather similar to Eq.~\eqref{eq:adjoint_master_appendix} and Eqs.~\eqref{eq:chi_mode1} and~\eqref{eq:chi_mode2}, respectively, at a first glance. However, there are significant conceptual differences.

The central difference is that the quantized quasi-normal modes are not based on the nonperturbative pseudomodes idea, since the approach does not attempt to choose a basis which is optimally Markovian \cite{Lentrodt2023}. By the latter feature, we mean to absorb all non-Markovian effects of the quantum system's dynamics with these effective discrete modes, such that the whole resulting extended system undergoes a Markovian dynamics, dictated by a master equation. 

While appealing from the classical resonance theory perspective, this approach necessarily leads to a non-optimal basis, relying on a Markov approximation \cite{Franke2019, Franke2019_supplement}, and frequency cutoffs \cite{Ren2020} in the derivation of the master equation. The method has so far mainly been benchmarked using Purcell factors for single and two-mode cavities \cite{Franke2019,Franke2020a,Ren2020,Fuchs2024_arxiv}, that is, for isolated resonances as they appear in good cavities.

However, the quantized quasi-normal modes approach also has one distinctive advantage over the standard pseudomodes: It incorporates the position degree of freedom in the framework. As described in the introduction Sec.~\ref{sec:intro}, this feature is also a main motivation of the gPMs developed here, which unites the two concepts.

Consequently, we can compare our approach to the quantized quasi-normal modes again by verifying if the matching condition is fulfilled. In the following, we show how the quantized quasi-normal mode expansion for the spectral density is computed. The method mainly follows \cite{Franke2019} adapted to one dimension \cite{Franke2023}. We note that, to obtain reasonable results, a different cutoff function compared to \cite{Ren2020} was used.

In the quantized QNM framework \cite{Franke2019}, the spectral density of the one-dimensional cavity example considered in Sec.~\ref{sec:applications} can be written with a similar construction as in Eq.~\eqref{eq:correlator_global_pseudomode}, as
\begin{align}
    &\Im \tilde C^\mathrm{qQNM}_{\mathrm{adv}}(x,x, \omega)\nonumber\\&= \frac{\hbar}{2\epsilon_0}\Im\left[\sum_{\lambda\lambda'}\sqrt{\omega_\lambda\omega_{\lambda'}}\tilde {f}^{\rm s}_{\lambda}(x)(\mathcal X-\omega)_{\lambda\lambda'}^{-1}\tilde { f}^{\rm s*}_{\lambda'}(x)\right],
\end{align}
where
\begin{align}
    \omega_\lambda &= \Re[\tilde\omega_\lambda],\\
    \mathcal X_{\lambda\lambda'} &= \sum_\eta S^{-1/2}_{\lambda\eta}\tilde\omega_\eta S^{1/2}_{\eta\lambda'},\\
    \tilde{f}_{\lambda}^{\rm s}(x) &= \sum_{\lambda'} S^{1/2}_{\lambda'\lambda}\sqrt{\frac{\omega_{\lambda'}}{\omega_\lambda}}\tilde{f}_{\lambda'}(x),
\end{align}
with $\tilde\omega_\lambda$ and $\tilde{f}_{\lambda'}(x)$ given by Eqs.~\eqref{eq:QNMs_1D} and~\eqref{eq:freqs_QNMs_1D}, respectively.

We therefore see that the objects $\mathcal X_{\lambda\lambda'}$ and $\tilde{f}_{\lambda}^{\rm s(x)}$ in the quantized QNM formalism are analogous to $\mathbb H$ and $\chi_{\lambda}(x)$ in the gPM approach [see Eqs.~\eqref{eq:correlator_global_pseudomode} and~\eqref{eq:chi_mode1}--\eqref{eq:coupling_ah}]. The difference lies in the matrix $S^{1/2}$: While, in our case, we use the Hermitization condition [Eqs.~\eqref{eq:hermiticity1} and~\eqref{eq:hermiticity2}] to construct suitable parameters, here $S^{1/2}$ is defined as the matrix square root\footnote{The matrix square root $S^{1/2}$, in this case, is defined as fulfilling $S = S^{1/2}S^{1/2}$.} of the following matrix, for a dispersionless medium:
\begin{align}
    S_{\lambda\lambda'} = \int_0^\infty\dd\omega\frac{2A_\lambda(\omega)A^*_{\lambda'}(\omega)}{\pi\sqrt{\omega_\lambda\omega_{\lambda'}}}S^{\mathrm{rad}}_{\lambda\lambda'}(\omega),\label{eq:franke_S_matrix}
\end{align}
where we introduced \cite{Ren2020}
\begin{align}
    A_\lambda(\omega) = \frac{\omega}{2(\tilde\omega_\lambda-\omega)}\mathrm{Rect}\left(\frac{\omega-\omega_\lambda}{\omega-\omega_\lambda+2\omega^{\mathrm{cut}}_\lambda}\right).\label{eq:A_cutoff}
\end{align}
Here, $\mathrm{Rect}(x)$ is a piecewise function which equals to $1$ for $-1/2<x<1/2$ and $0$ elsewhere, and $\omega^{\mathrm{cut}}_\lambda$ is a cutoff parameter which will be discussed in more detail below. In Eq.~\eqref{eq:franke_S_matrix}, we also used
\begin{align}
	S_{\lambda\lambda'}^\mathrm{rad}(\omega) = \frac{c^2}{2i\omega^2} \bigg[&\frac{\partial \tilde{F}_\lambda(x, \omega)}{\partial x}  \tilde{F}^*_{\lambda'}(x, \omega) \nonumber\\
	- &\tilde{F}_\lambda(x, \omega)  \frac{\partial\tilde{F}^*_{\lambda'}(x, \omega)}{\partial x}\bigg]\bigg|_{x=-\frac{L}{2}}^{x=\frac{L}{2}}, \label{eq::Srad_form}
\end{align}
where
\begin{align}
    \label{eq:reg_qnm}
    &\tilde{F}_\lambda(x, \omega) = \theta(x \in \mathcal{R})\tilde{f}_\lambda(x) \nonumber
    \\
    &+ \theta(x \notin \mathcal{R})\int_{-L/2}^{L/2} \dd x' \, G_\mathrm{B}(x, x', \omega) [n^2_{\rm R}-n^2_{\rm B}] \tilde{f}_\lambda(x) \,
\end{align}
are referred to as regularized QNMs \cite{Franke2019}. In Eq.~\eqref{eq:reg_qnm}, $\mathcal R=[-L/2, L/2]$ is the resonator's spatial extension and $G_{\mathrm{B}}(x,x',\omega)$ is the one-dimensional background Green's function, given by \cite{Kristensen2020}
\begin{align}
    G_\mathrm{B}(x, x', \omega) = \frac{c}{2i\omega n_\mathrm{B}} e^{i n_\mathrm{B}\omega|x-x'|/c}\,.\label{eq::Gbg}
\end{align}

The cutoff function in Eq.~\eqref{eq:A_cutoff} is introduced  based on \cite{Ren2020}, due to the fact that the integral in Eq.~\eqref{eq:franke_S_matrix} does not converge when integrated up to infinity. Note, however, that the cutoff function is modified compared to the one proposed in \cite{Ren2020}, to provide a reasonable approximation of the spectral density. As in \cite{Ren2020}, we chose the value $\omega^{\mathrm{cut}}_\lambda=-14\Im[\tilde\omega_\lambda]$ for the cutoff parameter.
\begin{figure}[t]
    \centering
    \includegraphics[width=\columnwidth]{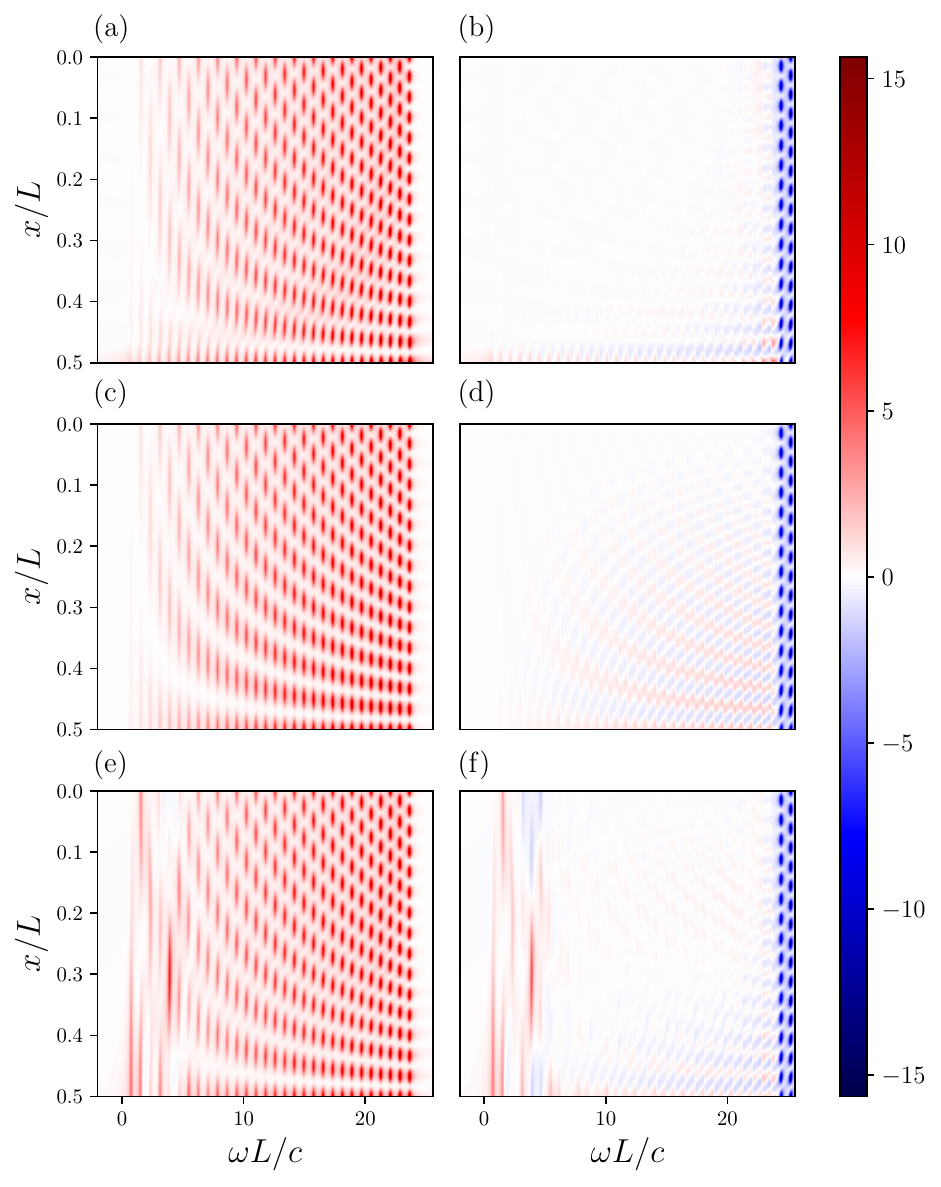}
    \caption{Comparison of spectral densities for the one-dimensional example cavity depicted in Fig.~\ref{fig:fig_1-1D_example}, choosing $n_{\mathrm{R}}=4$ and $n_{\mathrm{B}}=1$,  between different methods~\cite{WeitzelThesis2025}. In the left column, the figures (a), (c) and (e) show the spectral densities obtained from the gPMs approach, an alternative ``naive'' guess, and the quantized quasi-normal modes developed in \cite{Franke2019}, respectively. To compare with the exact spectral density, the right column shows the difference between the aforementioned spectral densities and the exact case. The colorbar is in units of $\hbar/(\epsilon_0 L)$. In all calculations, we used $M=30$ terms in the expansion.}
    \label{fig:fig_6-nR_4_2d_xw_comparison_Hermitization_naive_franke}
\end{figure}

Fig.~\ref{fig:fig_6-nR_4_2d_xw_comparison_Hermitization_naive_franke} shows the resulting comparison of the spectral density. The gPMs case, already shown in Fig.~\ref{fig:fig_4-spectral_density_plot}, is repeated for ease of comparison. We see that, while the quantized quasi-normal mode approach also performs well for intermediate frequencies, the deviations are larger than in the gPMs approach throughout, as expected from the Markov approximation. In addition, there are even larger deviations at low frequencies. These results confirm that the gPMs formalism provides a better approximation for the spectral density.

In addition to the quantized QNMs, Fig.~\ref{fig:fig_6-nR_4_2d_xw_comparison_Hermitization_naive_franke} compares a further alternative approach, which we refer to as ``naive'' Hermitization. From one perspective, our prescription can also be understood as trying to manipulate a resonance pole expansion [see Eq.~\eqref{eq:diag_propagator}] such that the numerators are compatible with a Hermitian field, as discussed in Sec.~\ref{sec:from_pseudomodes_to_pole}. The importance of this requirement was already pointed out in Ref.~\cite{Garraway1997a,SanchezBarquilla2022}, where it has been called an ``undiagonalization''.

We define the naive Hermitization as guessing the matrix solution of Eq.~\eqref{eq:hermiticity1} simply as
\begin{align}    
    T_{\mu\nu}\approx|\tilde\omega_\mu|\delta_{\mu\nu},\label{eq:naive_T}
\end{align}
that is, without taking into account the non-trivial properties of the quasi-normal modes [see Appendix~\ref{appendix:qnm_properties}] or the constraint of Eq.~\eqref{eq:hermiticity2}. With this --- at best highly approximate --- choice of the transformation matrix, one could again follow the steps outlined in Eqs.~(\ref{eq:hermiticity2} --~\ref{eq:coupling_ah}) to deduce an alternative set of modes and coupling parameters for the gPM ansatz [see Sec.~\ref{sec:gpm_setup}], avoiding the more involved approximation scheme introduced in Sec.~\ref{sec::gPM_approxSolHerm}. Effectively, this matrix then leads to the residues in the pole expansion Eq.~\eqref{eq:meromorphic_approx} as being approximated by
\begin{align}
    \sqrt{\tilde\omega_\mu}\tilde{\vb f}_{\mu}(\vb r)\approx \sqrt{\tilde\omega_\mu}^*\tilde{\vb f}^*_{\mu}(\vb r) \,.
\end{align}
More specifically, Eq.~\eqref{eq:naive_T} implies that the residues in the pole expansion of Eq.~\eqref{eq:diag_propagator} are approximately real. This assumption is not true in general, but can be a good approximation for good cavities with isolated resonances \cite{Lentrodt2023}. As shown in Fig.~\ref{fig:fig_6-nR_4_2d_xw_comparison_Hermitization_naive_franke}, this naive Hermitization yields a relatively good approximation for the spectral density, to the extent that the magnitude of the error shown in the corresponding differences plot is generally much smaller than the values the spectral density takes. However, we observe that the gPM formalism presented in Sec.~\ref{sec:gpm_theory} still leads to a better approximation, since the corresponding plot of differences depict an even wider area --- in Fig.~\ref{fig:fig_6-nR_4_2d_xw_comparison_Hermitization_naive_franke}, the region in white --- with an even smaller error, compared to the naive approach. Interestingly, the naive Hermitization already performs better than the quantized quasi-normal modes over a large range of the spectrum, illustrating the importance of the global pseudomode concept.

\bibliographystyle{myprsty}
\bibliography{library}

\end{document}